\documentclass{aa}  
\usepackage{xcolor}
\usepackage{graphicx}
\usepackage{txfonts}
\usepackage{lipsum}
\usepackage{subcaption} 
\usepackage{hyperref}
\usepackage{orcidlink}                                
\usepackage{lscape} 
\usepackage{colortbl}
\usepackage{booktabs} 
\usepackage{placeins}           
                                
\begin{document}
\nolinenumbers

\newcommand{\AK}[1]{\textbf{\textcolor{red}{(AK: #1)}}}
\newcommand{\SB}[1]{\textbf{\textcolor{blue}{(SB: #1)}}}
\newcommand{\tf}{\ensuremath{\mathrm{t_{50}}} }
\newcommand{\tq}{\ensuremath{\mathrm{t_{90}}} }
\newcommand{\tb}{\ensuremath{\mathrm{t_{0}}} }

\newcommand{\teff}{\ensuremath{\mathrm{t_{eff}}} }

\newcommand{\dhi}{\ensuremath{\mathrm{\delta_{HI}\,}}}
\newcommand{\dn}{\ensuremath{\mathrm{\delta_{\#}\,}}}
\newcommand{\ms}{\ensuremath{\mathrm{M_\star}} }
\newcommand{\msun}{\ensuremath{\mathrm{M_\odot}} }
\newcommand{\nd}{Na\ \textsc{D} }

   \title{Developing Environments: Formation and Quenching at Cosmic Noon Traced by the Large-scale Distribution of Gas and Galaxies}


   \titlerunning{Developing Environments}

\author{Amir H. Khoram\orcidlink{0009-0009-6563-282X}\inst{1,2}\corrauth{amirhossein.khoram@unibo.it}
        \and Sirio Belli\orcidlink{0000-0002-5615-6018}\inst{1}
        \and Trevor Mendel\orcidlink{0000-0002-6327-9147}\inst{3,4}
        \and Letizia Bugiani\orcidlink{0009-0008-7295-3288}\inst{1,2}
        \and Matteo Sapori\orcidlink{0009-0008-3970-4765}\inst{1,2}
        \and Andrew B. Newman\orcidlink{0000-0001-7769-8660}\inst{5}
        \and Hossein Hatamnia\orcidlink{0009-0007-3673-4523}\inst{5,6}
        \and Finn Giddings \orcidlink{0009-0003-2158-1246}\inst{7}
        \and Rebecca L. Davies\orcidlink{0000-0002-3324-4824}\inst{8,9}
        \and Olga Cucciati\orcidlink{0000-0002-9336-7551}\inst{2}
        \and Bianca Poggianti\orcidlink{0000-0001-8751-8360}\inst{10}
        \and Amirnezam Amiri \orcidlink{0000-0002-8553-1964}\inst{11,12}
        \and Minjung Park\orcidlink{0000-0002-8435-9402}\inst{13,14}
        \and Anita Zanella\orcidlink{0000-0001-8600-7008}\inst{2}
        \and Tianmu Gao\orcidlink{0000-0002-1158-6372}\inst{3,4}
        \and Gabriel Maheson\orcidlink{0009-0005-8978-8598}\inst{1}
        \and Stefano Sotira\inst{1}
        \and Caterina Liboni\orcidlink{0009-0004-2587-3577}\inst{1,2}
        \and Giulia Papini\orcidlink{0000-0002-5810-318X}\inst{2}}

   \institute{Dipartimento di Fisica e Astronomia, Università di Bologna, Via Gobetti 93/2, I-40129, Bologna, Italy 
   \and INAF, Astrophysics and Space Science Observatory Bologna, Via P. Gobetti 93/3, I-40129 Bologna, Italy 
   \and Research School of Astronomy and Astrophysics, Australian National University, Weston Creek, ACT 2611, Australia 
   \and ARC Centre of Excellence for All Sky Astrophysics in 3 Dimensions (ASTRO 3D), Canberra, ACT 2611, Australia  
   \and The Observatories of the Carnegie Institution for Science, 813 Santa Barbara St., Pasadena, CA 91101, USA
  \and Department of Physics and Astronomy, University of California, Riverside, 900 University Avenue, Riverside, CA 92521, USA
  \and Institute for Astronomy, University of Hawaii, 2680 Woodlawn Drive, Honolulu, HI 96822, USA
  \and ARC Centre of Excellence for All Sky Astrophysics in 3 Dimensions (ASTRO 3D)
   \and Centre for Astrophysics and Supercomputing, Swinburne University of Technology, Hawthorn, Victoria, Australia
   \and INAF – Osservatorio Astronomico di Padova, vicolo dell’Osservatorio 5, I-35122 Padova, Italy
   \and School of Astronomy, Institute for Research in Fundamental Sciences (IPM), Tehran, P.O. Box 19395-5531, Iran \and Department of Physics, University of Arkansas, 226 Physics Building, 825 West Dickson Street, Fayetteville, AR 72701, USA
   \and Kavli Institute for Cosmology, University of Cambridge, Madingley Road, Cambridge CB3 0HA, UK
   \and Cavendish Laboratory, University of Cambridge, 19 J. J. Thomson Ave., Cambridge CB3 0HE, UK}

   \date{Received September 30, 20XX}

  \abstract
   {Environmental density correlates with galaxy properties in the local Universe, but how and when these trends emerge remains unclear: at cosmic noon, large-scale structures are still assembling and processes such as galaxy quenching are proceeding rapidly for reasons that are still debated. In this work, we investigate how the formation and quenching histories of massive galaxies at $1.7<z<3.5$ depend on their surrounding environment, using two independent and complementary tracers of the environment: the galaxy number density (\dn) measured from COSMOS-Web and the distribution of intergalactic neutral hydrogen (\dhi, or cool gas)) measured from LATIS Ly$\alpha$ forest tomography maps. We combine 139 galaxies from the Blue Jay JWST/NIRSpec survey with 57 galaxies with prism spectroscopy from the DAWN JWST Archive, all in the COSMOS field. Stellar masses, star formation histories, dust attenuation, and Na\,{\sc d} neutral gas outflow rates are derived from joint \texttt{Prospector} fits to spectroscopy and photometry. Massive quenched galaxies show a clear environmental trend, with their formation-to-quenching interval increasing toward higher densities. AGN and Na\,{\sc d} neutral outflow signatures are confined to $\delta_{\#}<0.5$ and to regions of elevated Ly$\alpha$ absoption ($\delta_{\rm HI}>1$), with mostly massive quiescent galaxies displaying both simultaneously. Roughly $65\%$ of the sample follows a correlation between the two environmental tracers; among the outliers we identify two interesting regions where the two tracers are not in agreement. One is a galaxy-rich, cool gas-poor region ($\dn > 2$, $\dhi < 1$) which is coincident with one of the most massive cores in the Hyperion protocluster; the lack of neutral hydrogen is plausibly explained by the fact that the gas is fully in the ionized phase. Quiescent galaxies in this region formed earlier and lack substantial AGN activity and neutral outflow signature. The second region is galaxy-poor, gas-rich ($\dn \sim 1$, $\dhi > 1$) and is located in 60 cMpc radius from densest Hyperion cores. Here, quiescent galaxies appear to have quenched rapidly and present a remarkably high incidence of AGN activity and neutral gas outflows. This \emph{developing environment} is distinct from both mature protoclusters and field, and likely coincides with massive halos which are actively assembling, so that the galaxy count is still relatively low. We conclude that the environment at cosmic noon cannot be captured by galaxy counts alone. The joint analysis of galaxy density and cool gas distribution in this work reveals distinctions that crucial for understanding how environment regulates gas accretion, star formation, and ultimately galaxy quenching.

   }


   \maketitle

\section{Introduction}

A central aim of extragalactic research is to establish the relative contribution of internally versus externally driven physical processes throughout the evolution of galaxies. In the nearby universe, environmental density correlates with several galaxy properties such as their star-formation, stellar mass, and metallicity \citep[e.g.,][]{2003Gomez,2004Kauffmann,2006Poggianti,2025Khoram}, with these trends established as early as $z \sim 1$ \citep[e.g.,][]{2012Muzzin,2016Darvish}. At higher redshifts, however, large-scale structures are still actively assembling, shaped by continuous growth from the reionization era through Cosmic Noon \citep[e.g.,][]{2015Muldrew}, so the environmental indicators used in the local Universe are no longer reliable tracers of galaxy evolutionary stage. Tracing the redshift evolution of these environmental trends is therefore essential for identifying their physical origins, which likely reflect a combination of density-sensitive processes such as ram pressure stripping, starvation, and tidal interactions, and differences in halo assembly history \citep[e.g.,][]{2026Roberts,2026Xu,2026Zhang}. 

Among the most important processes governing galaxy evolution is the formation and quenching of massive galaxies. Massive quiescent galaxies detected in the early Universe seem to have experienced a swift halt in star formation, although the underlying processes driving this transformation are still widely debated \citep[e.g.,][]{2018NMan&Belli,2026Whitaker}. The Cosmic Noon epoch — marked by elevated rates of star formation, galactic quiescence, and halo development — offers a valuable opportunity to examine these interconnected processes \citep{2009Dekel,2014Madau,2019Behroozi}. Although feedback from active galactic nuclei (AGN) is commonly cited as the dominant mechanism responsible for quenching star formation  \citep{2005DiMatteo,2015Somerville, 2017Beckmann, 2024Belli, 2025Bugiani,2026Davies}, alternative investigations emphasize the considerable contributions of the environment, such as strangulation, gas exhaustion,  and dynamical disturbances triggered by galaxy mergers \citep{2024Ellison,2025Suess,2026khoram}.

The physical mechanisms responsible for the rapid buildup and subsequent early quenching of galaxies in various environments remain largely unclear. A key obstacle lies in the difficulty of detecting protoclusters, which at $z > 2$ are neither fully collapsed nor virialized, extending across tens of comoving megaparsecs \citep{2013Chiang, 2015Muldrew}. Unlike mature clusters, these structures may not have yet developed a hot intracluster medium, rendering standard cluster-finding techniques including extended X-ray emission and the Sunyaev-Zel'dovich effect largely ineffective. Identifying protoclusters therefore hinges on the statistical detection of galaxy overdensities. Despite significant progress in the pre-JWST "era" with ground- and space-based instruments  \citep[e.g.,][]{2016Laigle,2017Fossati,2022Weaver}, measurements of galaxy environments at high redshift remained limited by relatively large photometric-redshift uncertainties, shallow imaging depths, or high stellar-mass completeness limits. These limitations made environmental studies beyond $z \sim 2$ increasingly uncertain, with surveys primarily sensitive to the most massive and luminous galaxies \citep[e.g.,][]{2025Baxter}. Consequently, tracing the full diversity of environments and the emergence of large-scale structure during cosmic noon remained challenging.

The advent of JWST has significantly improved these measurements through its unprecedented depth and near-infrared imaging and grism spectroscopy capabilities, enabling more accurate photometric redshifts and the detection of substantially fainter galaxy populations. Recent studies using JWST data have begun reconstructing the large-scale structure and environmental density field across several deep extragalactic surveys \citep[e.g.,][]{2023Casey,2026Hatamnia}. In particular, the COSMOS-Web survey \citep{2023Casey} has enabled environmental measurements over a contiguous $\sim0.54\,\mathrm{deg}^2$ field with greatly improved redshift precision and mass completeness, allowing the identification of overdense structures and the study of environment-driven galaxy evolution out to $z\sim7$.

Advances in photometric redshifts and galaxy number counts have been complemented by large spectroscopic programs which have constructed three-dimensional maps of the cool gas distribution. By targeting Ly$\alpha$ absorption imprinted on the spectra of background Lyman-break galaxies (LBGs) and quasars, these maps trace the spatial distribution of neutral hydrogen (HI) with high fidelity, providing a view of the intergalactic medium (IGM) that is independent of galaxy counts \citep[e.g.,][]{2019Horowitz,2020Newman,2021Ata}. Crucially, the two tracers are not necessarily interchangeable: comparisons at $z\sim2$--3 reveal a relation far more complex than a one-to-one correspondence \citep[e.g.,][]{2021Momose,2022Newman,2023Dong}, including configurations in which galaxy-rich regions appear HI-poor and vice versa. Such mismatches have so far been mostly interpreted as a consequence of incomplete galaxy censuses, in which galaxies in these regions are missed owing to their intrinsic properties, such as low stellar mass or strong, wavelength-dependent obscuration that dims their observed luminosity.

In this work, we leverage both probes of cosmic structure available in the well-studied COSMOS field: the galaxy number densities measured from COSMOS-Web \citep{2026Hatamnia} and the neutral hydrogen maps derived by the LATIS survey \citep{2020Newman}.
Within the COSMOS field, the most prominent structure at cosmic noon is the Hyperion proto-supercluster \citep{2018Cucciati}, an extended overdensity at $z\simeq2.45$ spanning tens of comoving megaparsecs and comprising several distinct density peaks. Its multi-component morphology is characteristic of a structure caught mid-assembly, with several cores expected to differ in mass, dynamical state, and gas content; constrained simulations predict that these peaks will evolve into separate clusters by the present day \citep{2022Ata}. Hyperion therefore offers a rare opportunity to examine environmental effects within a structure whose subcomponents span a range of evolutionary stages, provided that results obtained for one core are not generalized to the whole structure.

Our goal is to address two of the most actively debated questions in galaxy evolution: what drives the rapid formation and quenching of massive galaxies, and what governs the persistence of quenching once it is established. In addition to the environmental measurements, we make use of recent JWST/NIRSpec data with medium spectral resolution from the Blue Jay survey \citep{2026Belli}, which we augment with a sample of public low-resolution prism spectra. We are thus able to link IGM properties to the stellar mass, AGN activity, outflow signatures, and SFH (or equivalently mass assembly history) of individual galaxies. Non-parametric SFHs now allow formation and quenching epochs to be dated for individual systems at cosmic noon \citep[e.g.,][]{2019Lejaa, 2024Park}, while JWST spectroscopy has begun to directly reveal the feedback acting during the transition itself \citep[e.g.,][]{2024Davies,2024Belli,2024DEugenio}. Combining these internal diagnostics with independent environmental measurements allows us to compare the contributions of the two quenching channels. Our aim is not only to build a more complete picture of the interplay between galaxies and their large-scale environment, but also to shed light on the development of the environment itself.

This paper is organized as follows. In Sect.~\ref{Sec:Data}, we present our spectroscopic sample, while in Sect.~\ref{Sec:Environment Measurements} we introduce the independent environmental measurements used throughout this work. In Sect.~\ref{Sec:Environment and Formation/Quenching}, we investigate how galaxy formation and quenching depend on environment, followed by Sect.~\ref{Sec:Neutral gas and galaxy number density}, where we explore neutral hydrogen (\textsc{HI}) as an independent tracer of environment. We discuss the implications of our results in Sect.~\ref{Sec:Discussion} followed by the summery of our main findings. Throughout this paper, we adopt the \cite{2020Planck} cosmological parameters.

\section{Data}\label{Sec:Data}

\begin{figure}
    \centering
    \includegraphics[width=\linewidth]{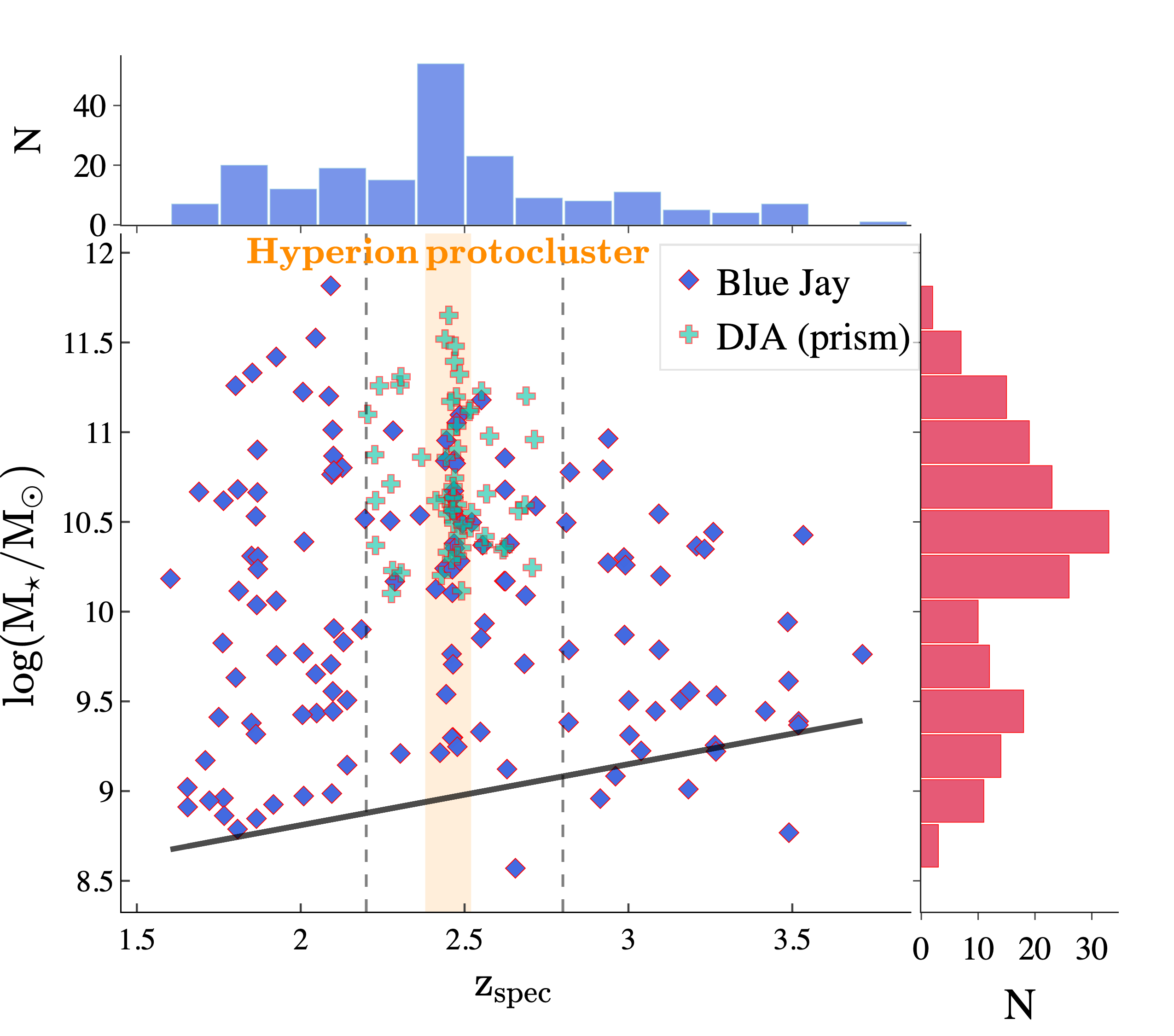}
    \caption{Spectroscopic redshift-\ms  plane of the Blue Jay and DJA sample used in this study. The stellar mass and redshift distributions are shown in the right and top histograms, respectively. The solid black line represents the stellar mass completeness limit of the sample. The shaded region indicates the redshift range occupied by the Hyperion protocluster. Moreover, the dashed lines represent the redshift range where both types environmental measure employed in this work are available (see Sect.\ref{Sec:Environment Measurements}).}
    \label{fig:SampleDist}
\end{figure}
 
\subsection{The Blue Jay sample}\label{Sec:The BlueJay sample}

The Blue Jay survey comprises 153 galaxies in the COSMOS field observed with JWST/NIRSpec as part of Cycle~1 program GO~1810 (PI: Belli). After excluding four $z\sim6$ filler targets and ten galaxies for which no spectroscopic redshift could be determined owing to the absence of identifiable spectral features, the final sample contains 139 galaxies spanning $8.8<\log(M_{\star}/M_{\odot})<11.5$ at $1.7<z<3.5$, as illustrated in Fig.~\ref{fig:SampleDist}. Medium-resolution NIRSpec MSA spectroscopy (G140M, G235M, and G395M) provides rest-frame coverage from $3700$~\AA{} to $1.2~\mu$m. The spectra were reduced using a modified JWST Science Calibration Pipeline, including master-sky subtraction derived from empty shutters. Parallel JWST/NIRCam imaging in eight filters was supplemented with public PRIMER imaging, and with archival HST/ACS+WFC3 photometry to extend the wavelength coverage into the UV. Full survey details are presented in \citet{2026Belli}.

\begin{figure}
    \centering
    \includegraphics[width=\linewidth]{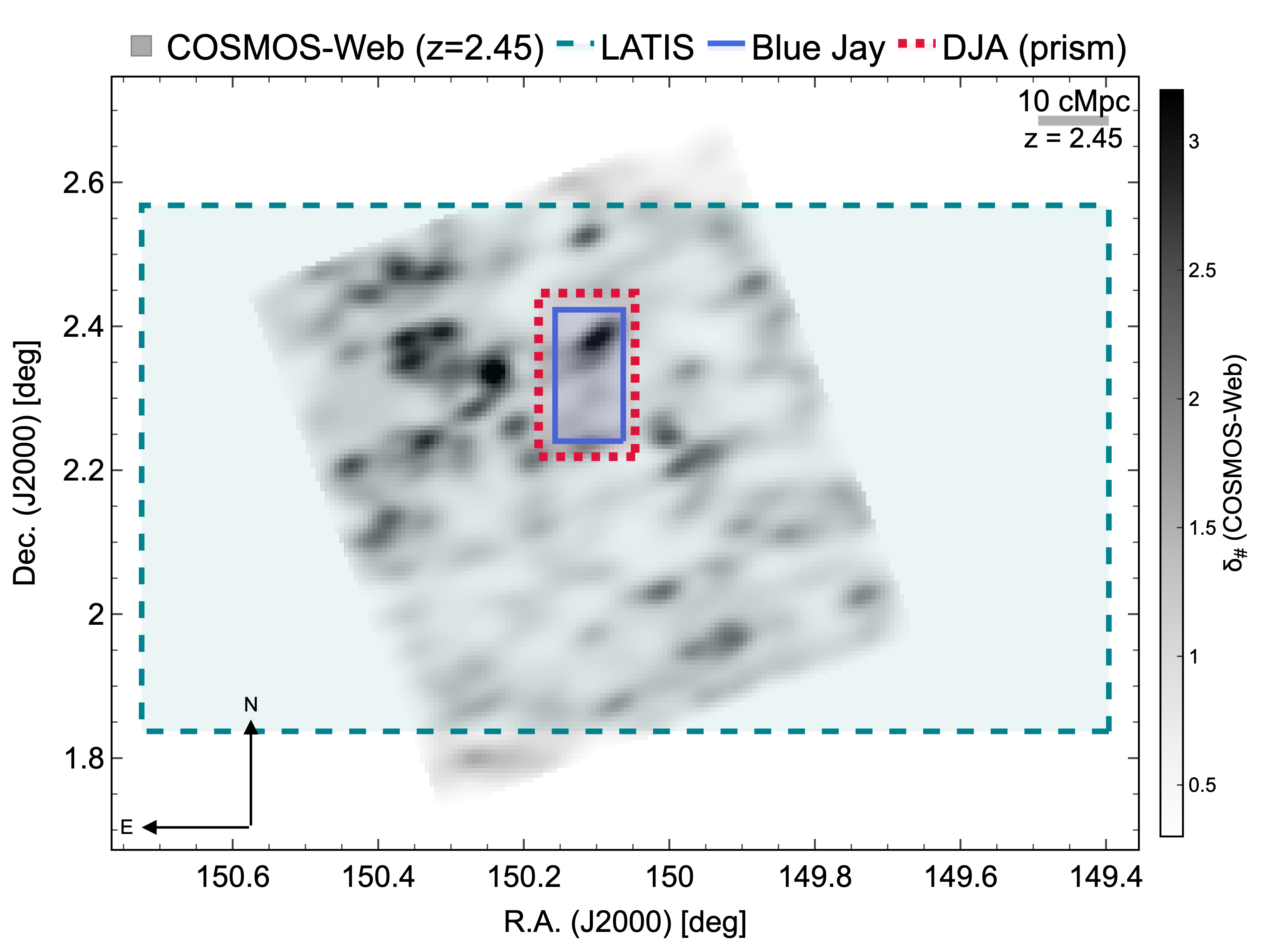}
    \caption{Sky footprints of the datasets used in this work, overlaid on the COSMOS-Web galaxy overdensity map at $z=2.45$ in the vicinity of Hyperion protocluster (grayscale, $\delta_{\#}$; Sect.~\ref{Sec:COSMOS-WEB}). The dashed teal rectangle marks the area covered by the LATIS IGM tomography, within which \dhi\ is measured (see Sect.~\ref{Sec:LATIS}). The solid blue and dotted red rectangles show the regions covered by the Blue Jay and DJA (prism) spectroscopic samples, respectively; both lie well inside the LATIS and COSMOS-Web coverage, so every galaxy in our sample has a measurement of both environment tracers. The scale bar in the top right corner indicates 10 cMpc at $z=2.45$, and the compass in the bottom left shows the orientation on the sky.
\label{fig:footprints}}
\end{figure}

We infer the stellar population properties using \texttt{Prospector} \citep{20121Johnson}, jointly fitting the spectra and photometry following the methodology of Bugiani et al. (in prep.). As described in detail by \citep{2026Belli} and Bugiani et al. (in prep.), \texttt{Prospector} was applied to the ``box'' photometry, extracted from the same spatial region intercepted by the NIRSpec MSA slitlets and used for the 1D spectral extraction. The modelling adopts FSPS \citep{2010Conroy}, the MIST isochrones \citep{2016Choi}, the MILES spectral library \citep{2012miles,2016miles}, a Chabrier IMF \citep{2003Chabrier}, and a non-parametric SFH \citep{2019Lejaa,2019Lejab}. Stellar masses and star-formation rates (SFRs) are derived from the best-fit SFHs, with the SFR averaged over the past 30 Myr. We adopt a non-parametric SFH with 14 time bins in lookback time. The youngest bin spans $[0, 30\,\mathrm{Myr}]$, the intermediate bins are logarithmically spaced out to $0.8\,t_{\rm univ}$, where $t_{\rm univ}$ is the age of the Universe at the galaxy's redshift, and the final two bins are fixed to $[0.8\,t_{\rm univ}, 0.9\,t_{\rm univ}]$ and $[0.9\,t_{\rm univ}, 1.0\,t_{\rm univ}]$, providing sensitivity to early mass assembly. Dust attenuation is modelled with a two-component prescription: a diffuse component following the \citet{2013Kriek} attenuation law, and a birth-cloud component affecting stars younger than 10 Myr. Emission-line fluxes are measured by subtracting the best-fit stellar continuum and fitting Gaussian profiles with \texttt{emcee} \citep{2013Foreman}. All lines share a common redshift and velocity dispersion, and the fit accounts for the wavelength-dependent NIRSpec spectral resolution. 

In our sample, AGNs are identified via either BPT diagram \citep{1981BPT} classification using emission line catalogs provided by \cite{2025Bugiani} and Bugiani et al. (in prep) or through X-ray detection \citep{2025Ito, 2026Belli}, although none of the X-ray-detected AGNs is a quiescent galaxy. For the former, we adopt the \citet{2003Kauffmann} demarcation lines in both the N\,{\sc ii} BPT and S\,{\sc ii} BPT diagrams.

In both galaxy samples, we classify galaxies as quenched when their SFR falls more than 1\,dex below the star-forming main sequence at their respective redshift, following \citet{2018Fang,2019Donnari}. We verify that adopting an alternative criterion based on $\mathrm{sSFR} < -10.5 \,\mathrm{yr^{-1}}$, following \cite{2019Lejaa}, yields an essentially identical massive quiescent galaxy sample, differing by only one additional galaxy. Additionally, only 10 of the 15 Blue Jay quiescent galaxies (see Tab.~\ref{tab:BJ_Tab}) are in common with the earlier Blue Jay samples of \citet{2024Park} and \citet{2025Bugiani}, owing to the improved data reduction and updated methodology described in \citet{2026Belli} and Bugiani et al.\ (in prep.).

Throughout this work, we classify galaxies as \emph{massive} when $\log(M_\star/M_\odot)>10$. Reassuringly, in both cases the identified quenched galaxies predominantly populate the massive end of the distribution.

\subsection{DJA prism}\label{Sec:DJA prism}

Using the DAWN JWST Archive (DJA)\footnote{\href{http://dawn-cph.github.io/dja/index.html}{http://dawn-cph.github.io/dja/index.html}}, we select 244 galaxies with prism spectroscopy in a field similar to the Blue Jay coverage (see Fig.~\ref{fig:footprints}) in order to enlarge our sample at $2.2<z<2.8$, the redshift range where we can quantify environments using both galaxy number counts and Ly$\alpha$ tomography (see Sects.~\ref{Sec:COSMOS-WEB} and~\ref{Sec:LATIS}). From these we retain 57 massive galaxies with $\log(M_\star/M_\odot)>10$, as illustrated in Fig.~\ref{fig:SampleDist}. In addition to the prism spectroscopy, our analysis incorporates ancillary photometry from the ASTRODEEP-JWST survey\footnote{\href{http://www.astrodeep.eu/}{http://www.astrodeep.eu/}}. These catalogs \citep{2024Merlin} combine measurements in 16 photometric bands, including reprocessed HST imaging, and cover a wavelength range of roughly 0.44--4.44\,$\mu$m.

We simultaneously model the extracted prism spectra and photometric data using \texttt{Prospector} \citep{20121Johnson}, adopting the same stellar population synthesis library, isochrones, spectral library, IMF, and non-parametric SFH prior as used for the Blue Jay sample (see Sect.~\ref{Sec:The BlueJay sample}), with the spectral fitting parameters adjusted to the lower resolution of the prism. Unlike the default Blue Jay fits, which are restricted to the rest-frame 3610--6700\,\AA\ window to avoid the calibration polynomial distorting the 4000\,\AA\ break, we fit the entire prism spectrum jointly with the photometry. We also treat the redshift as a free parameter with a tight Gaussian prior (\(\sigma = 0.05\)) centered on the DJA redshift, to account for the larger redshift uncertainties inherent to low-resolution prism spectroscopy.

While the classic N\,{\sc ii}--BPT diagram cannot be applied to our prism spectra, since the low spectral resolution blends [\ion{N}\,{\sc ii}]$\lambda6584$ with H$\alpha$, two galaxies are classified as AGN based on their excess [\ion{O}{I}]$\lambda6300$ flux, following the diagnostic of \citet{2006Kewley}. Because H$\alpha$ cannot be measured directly, we estimate it from the observed H$\beta$ flux assuming the recombination value (i.e., 2.86) and no dust attenuation; since dust preferentially suppresses H$\beta$ relative to H$\alpha$, this yields a conservative lower limit on the true H$\alpha$ flux.

\section{Environment Measurements}\label{Sec:Environment Measurements}
In this work, we employ two distinct measurements of environment at cosmic noon: traditional galaxy number-density maps and Ly$\alpha$ forest tomographic maps. While number-density maps trace the spatial distribution of galaxies, Ly$\alpha$ tomography probes the distribution of cool gas, specifically neutral hydrogen (\textsc{HI}), within the large-scale structure. As neutral gas represents a fundamental component of galaxy growth and evolution, mapping the \textsc{HI} distribution alongside galaxy overdensities provides a more complete view of the environmental conditions in which galaxies evolve. However, Ly$\alpha$ forest tomography is observationally expensive, requiring large samples of high-quality background spectra with sufficient signal-to-noise ratio and spectral resolution, typically obtained from thousands of sources over wide fields. As a result, only a limited number of such reconstructions have been achieved to date, primarily in a few well-studied deep fields and over restricted redshift ranges \citep[e.g.,][]{2014Lee,2019Horowitz,2020Newman}.

\begin{figure*}
\begin{minipage}{0.73\textwidth}
    \centering
    \includegraphics[width=13cm,height=1\textheight,keepaspectratio]{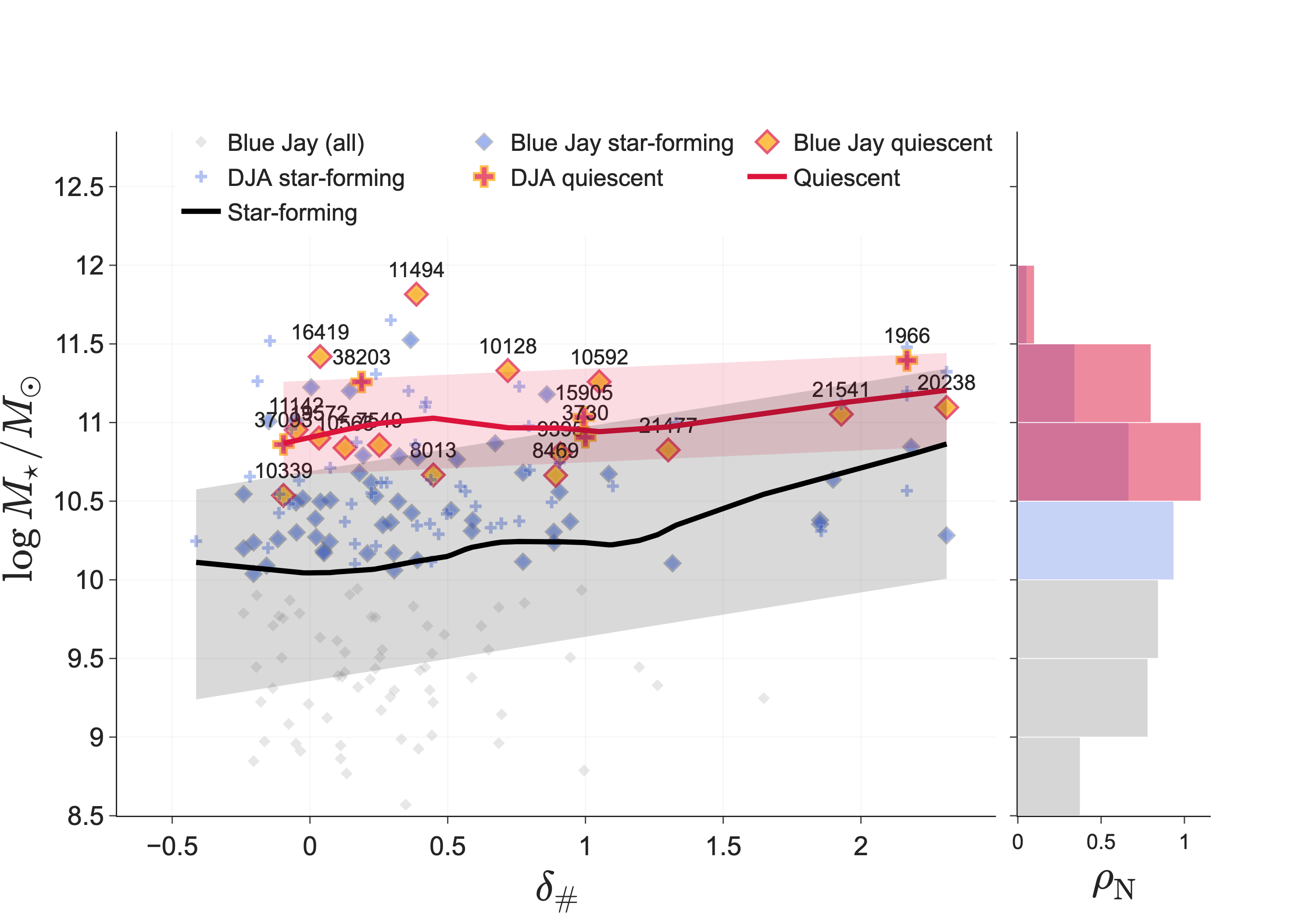}
\end{minipage}\hfill
\begin{minipage}{0.27\textwidth}
    \caption{The stellar mass distribution of the full sample as a function of galaxy number density (see Sect.~\ref{Sec:COSMOS-WEB}). The solid red and black curves show the LOWESS (Locally Weighted Scatterplot Smoothing) fits to the massive quiescent and star--forming galaxy populations, respectively. The shaded regions represent the $1\sigma$ scatter of each population around the corresponding linear fit. The right panel shows the stellar mass distributions of the three galaxy classes considered in this work: massive quiescent galaxies, massive star-forming galaxies, and non-massive star--forming galaxies.}
    \label{fig:mass delta}
\end{minipage}
\end{figure*}

Combining these two environmental tracers is particularly powerful at high redshift, where cosmic structures are still assembling and environmental processes are actively emerging. The joint analysis of galaxy density and cool gas distributions enables us to investigate the interplay between environment, galaxy formation, and evolutionary pathways, including the physical mechanisms driving star formation bursts and quenching during cosmic noon.

\subsection{COSMOS-WEB Number Density Map ($\mathrm{\delta_\#}$)}\label{Sec:COSMOS-WEB}
We use the number density maps from \cite{2026Hatamnia}, which reconstruct the galaxy density field across the footprint of the COSMOS-Web survey out to $z \sim 9.5$, complete down to $\log(M_\star/M_\odot) > 8.0$. The reconstruction is based on a weighted kernel density estimation (wKDE) applied to $\sim160{,}000$ galaxies with robust photometric redshifts derived from up to 30 photometric bands, including JWST imaging. The methodology, as well as the source selection and masking procedures, are described in detail in \cite{2026Hatamnia}. The density field is computed in fixed slices of width $35\,h^{-1}$ cMpc, resulting in 157 redshift slices across the full survey volume. Each galaxy is assigned a probabilistic weight in each slice according to its redshift PDF, and the density maps are generated using wKDE while accounting for edge effects, adaptive bandwidths, and regions masked by bright stars. This framework enables the characterization of galaxy environments over a wide range of redshift and density regimes up to $z \sim 7$.

For each galaxy in our sample, we assign an overdensity value, $\delta_\#$, by matching its sky coordinates and redshift to the nearest density slice from \cite{2026Hatamnia}. The overdensity used in our work is defined as
\begin{equation}
\delta_\# =  \frac{\sigma_s(\chi_i)-\bar{\sigma}_s}{\bar{\sigma}_s}  \equiv  \delta ,
\label{eq:dN}
\end{equation}
where $\sigma_s(\chi_i)$ is the local surface number density at the galaxy position $\chi_i$, and $\bar{\sigma}_s$ is the mean surface number density within the corresponding redshift slice. Galaxies in the average density field will thus have  \dn$= 0$, while overdensities are characterized by \dn $> 1$. We verified that averaging over neighboring slices in redshift and projected spatial position does not significantly alter the assigned \dn values for individual galaxies.

The stellar mass distribution of our full sample is shown in Fig.~\ref{fig:mass delta} as a function of the galaxy number density measured from COSMOS-Web. Star-forming galaxies show an apparent positive correlation between stellar mass and galaxy number density, which could be interpreted as the well-established mass--environment relation. However, we refrain from drawing conclusions from this trend, given the relatively high mass-completeness limit ($\sim 10^{9}\msun$) of the sample and the small number of low- and intermediate-mass galaxies detected in the highest-density regions.

In contrast, the massive quiescent galaxies have an approximately uniform stellar mass distribution (right panel of Fig.~\ref{fig:mass delta}) and are distributed largely uniformly across environments. More specifically, we find no significant correlation between stellar mass and \dn (Spearman $\rho = +0.21$, $p = 0.31$), with a Theil--Sen slope of $+0.10$ ($95\%$ CI $[-0.11, +0.28]$) that limits any mass increase across the probed range. Stellar mass therefore does not drive any environmental differences within this population, and the trends discussed throughout this work cannot be attributed to a mass--environment effect.

\subsection{LATIS Ly$\mathrm{\alpha}$ Tomographic Map (\dhi)}\label{Sec:LATIS}

We use the tomographic density field reconstruction presented in LATIS \citep{2025Newman}, which maps the large-scale structure at $2.2<z<2.8$ using intergalactic neutral hydrogen (\ion{H}{i}) absorption seen against a dense sample of background sources. The reconstruction is based on $\sim4{,}000$ LBGs and quasars in all the Canada--France--Hawaii Telescope Legacy Survey fields, including COSMOS, which provide sightlines through a $\sim1.6$\,deg$^2$ field of view. The background sources reach typical depths of $i=24.8$\,mag, enabling a sightline spacing of order $\sim1$ physical Mpc at these redshifts. The density field is reconstructed on a three-dimensional grid by applying Wiener filtering to the Ly$\alpha$ forest flux field, with covariance models for intergalactic-medium fluctuations and observational noise. The tomographic maps have a redshift resolution of $\Delta z\sim0.005$. The final product is a smoothed Lya transmission field sampled in redshift slices across the COSMOS footprint.

The Ly$\alpha$ forest flux fluctuation field is defined as
\begin{equation}
\delta_F = \frac{F}{\langle F(z) \rangle} - 1,
\end{equation}
where $F$ is the transmitted flux fraction and $\langle F(z)\rangle$ is the mean transmission at redshift $z$, taken from measurements in quasar spectra \citep{2008Faucher}. Since $\delta_F$ encodes the strength of Ly$\alpha$ absorption, larger absorption corresponds to more negative values of $\delta_F$, indicating higher cool gas ($<10^4 \,$K) density. Conversely, weaker absorption results in higher values of $\delta_F$, corresponding to lower \textsc{HI} density. 

Following \cite{2020Newman}, flux minima are identified within Wiener-filtered maps smoothed using a Gaussian kernel of width $\sigma_{\rm kern}=4\,h^{-1}\,\mathrm{cMpc}$. Each map is then normalized by its standard deviation, $\sigma_{\rm map}$, yielding
\begin{equation}
\delta_{\mathrm{HI}} = -1 \times \frac{\delta_F}{\sigma_{\rm map}} 
\label{eq:DH}
\end{equation}
which we adopt as a tracer of the neutral hydrogen density field within each redshift slice. By construction, the mean of the field is set to zero, such that $\delta_{\rm HI}=0$ represents the average Ly$\alpha$ absorption in each redshift slice. Negative values of $\delta_{\rm HI}$ therefore trace underdense, weaker absorbing regions, while positive values correspond to overdense, stronger absorbed regions relative to the mean.

\begin{figure*}
    \centering
    \includegraphics[width=0.9\linewidth]{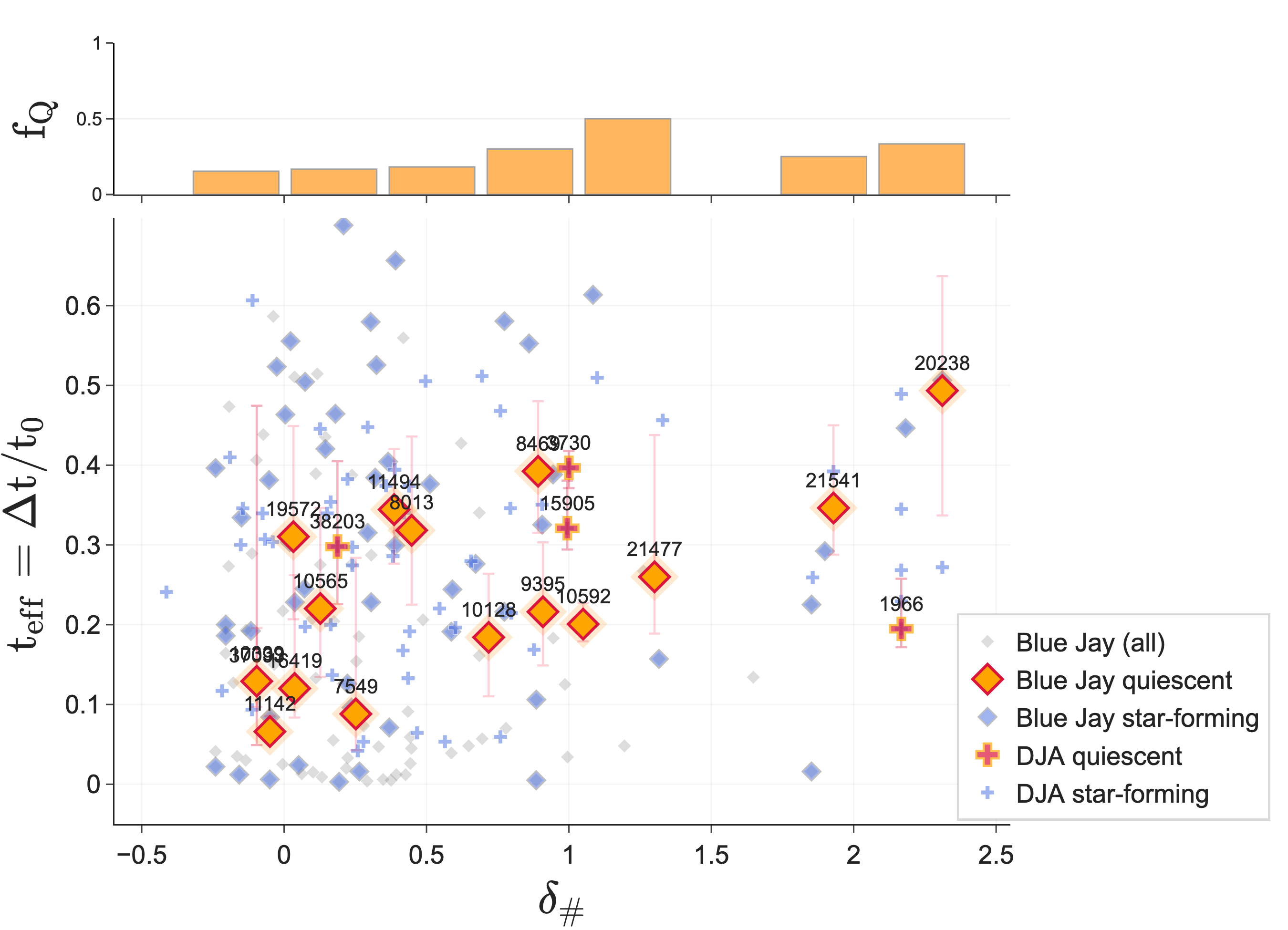}
\caption{The effective lifetime $\teff = \Delta t_{(50-90)}/\tb$ as a function of the local galaxy number density \dn. Grey diamonds show the full Blue Jay sample. Coloured symbols mark massive galaxies ($\log M_\star/M_\odot > 10$ ), split into quiescent (orange/red) and star-forming (blue) systems. Diamonds are Blue Jay galaxies and crosses are DJA prism galaxies; quiescent galaxies are labelled with their IDs. Error bars on the quiescent galaxies show the 1$\sigma$ standard deviation of the \teff\ posterior. The upper panel shows the quiescent fraction, $f_{\rm Q}$, among massive Blue Jay galaxies in bins of \dn. }
    \label{fig:t90t50_D}
\end{figure*}

Notably, throughout this work, we assign density values to each galaxy by projecting it onto the nearest redshift slice and extracting the corresponding spatial density from each map. We note that, since both density fields used in this work are already smoothed, averaging over neighboring galaxies in redshift and on the sky does not alter the resulting density values.

\section{The Role of Environment in the Formation and Quenching of Galaxies}\label{Sec:Environment and Formation/Quenching}

Using inferred stellar masses and SFHs from \texttt{Prospector}, we define two characteristic timescales, \tf\ and \tq, corresponding to the lookback time at which a galaxy assembled $\mathrm{50\%}$ and $\mathrm{90\%}$ of its final stellar mass, respectively. Throughout this work, we refer to \tf\ as the “formation time.” For quenched galaxies, \tq\ is interpreted as the “quenching time,” implying how long ago the galaxy started the quenching. These quantities provide a consistent framework for comparing mass assembly histories across different galaxy populations and environments.

Although the absolute values of \tf and \tq encode meaningful information about each galaxy's evolutionary stage, the broad redshift range spanned by our sample ($1.7 < z < 3.5$) complicates any direct, epoch-independent comparison of these timescales. A more robust approach is to normalize both quantities by the age of the universe at the redshift of observation, \tb, which largely removes redshift-dependent effects and instead reveals the intrinsic shape of the star formation history. This is achieved by considering the normalized quantities $\tf/\tb$ and $\tq/\tb$, or equivalently by defining an "effective lifetime," $\mathrm{t_{eff}} =(\tf - \tq)/\tb$, which is the fraction of cosmic time elapsed between a galaxy's 50\% and 90\% mass-assembly lookback times. For quenched galaxies, small values of this quantity indicate galaxies that quenched shortly after reaching half their final stellar mass, while larger values point to more gradual quenching pathways, in which most of the stellar mass was assembled early and quenching proceeded over a much longer timescale. Framed this way, the effective lifetime offers a redshift-independent metric for comparing formation and quenching histories consistently across the entire sample. Note that, \teff uncertainties are defined as the 16th--84th percentiles of the \teff\ posterior, with \teff\ computed for each of 1000 posterior samples of the SFH so that the covariance between $t_{50}$ and $t_{90}$ is included.

\subsection{Formation and Quenching }\label{formation and quenching}

In Fig.~\ref{fig:t90t50_D}, the distribution of \teff for star-forming and quiescent galaxies in the Blue Jay and DJA samples is shown as a function of galaxy number density \dn. The massive quiescent galaxies exhibit a clear trend with \dn, with \teff increasing toward higher-density environments. We find a positive correlation between \teff and \dn with Spearman $\rho = +0.6$, $p = 0.01$ from a permutation test and A Theil--Sen fit gives $\mathrm{d}t_{\rm eff}/\mathrm{d}\delta_{\#} = +0.1$, with a 95\% confidence interval of $[+0.1, +0.2]$, corresponding to an increase of $\approx 0.2$ in $t_{\rm eff}$ across the probed density range.

\begin{figure*}
    \centering
    \includegraphics[width=\linewidth]{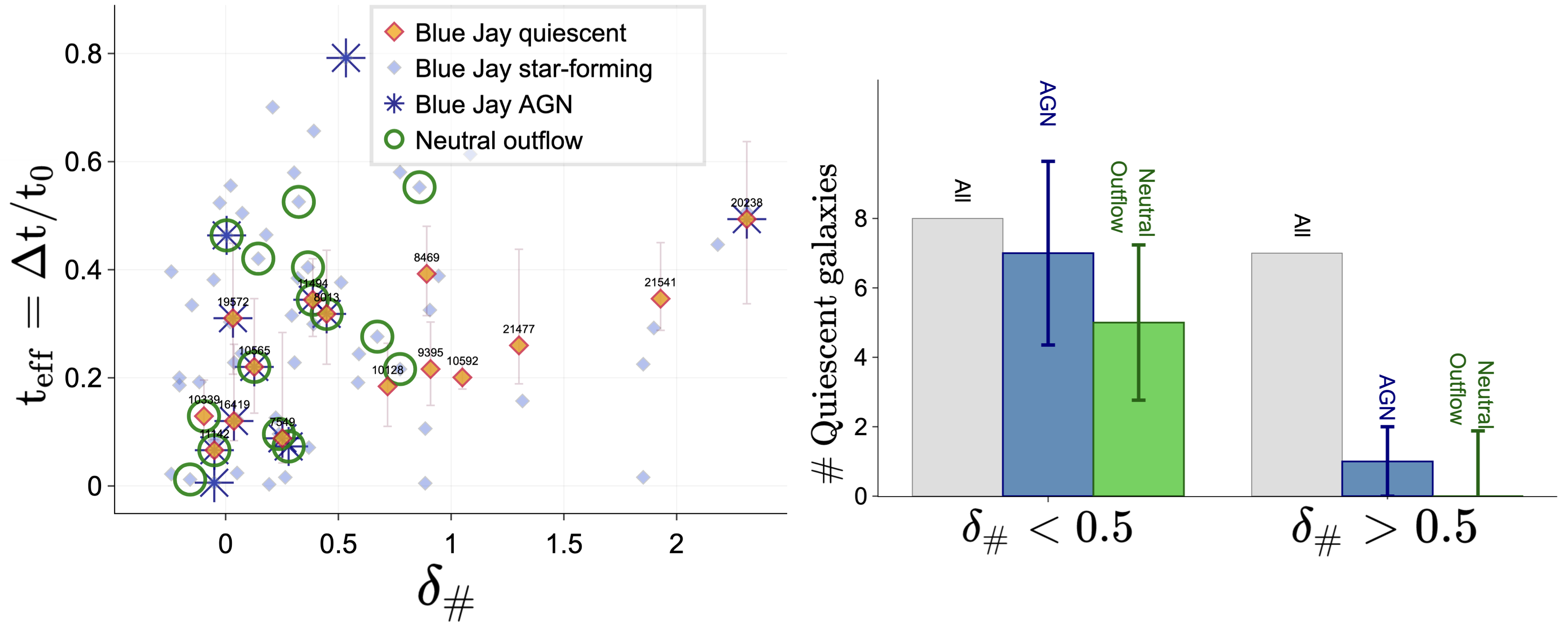}
    \caption{\textbf{Left panel:}  The effective lifetime $\teff = \Delta t_{({50}-{90})}/\tb$ as a function of local galaxy number density \dn spanning the full sample redshift range, $1.7<z<3.5$. Massive quenched galaxies and massive star-forming galaxies are shown in orange and blue, respectively. Error bars on the quiescent galaxies show the 1$\sigma$ standard deviation of the \teff\ posterior. Galaxies hosting AGN, identified either through the BPT diagram or X-ray detections, are marked with blue symbols, while galaxies exhibiting neutral outflows traced by \nd absorption are indicated with green symbols.  \textbf{Right panel:}  The bar chart shows the number of massive quiescent galaxies that host AGN or exhibiting neutral outflow signatures in their spectra, for two subsamples split by local galaxy number density. The error bars are derived assuming Poisson statistics and indicate the corresponding $2\sigma$ confidence intervals. Note that the \dn$=0.5$ threshold is chosen purely to delineate the region where the AGN are concentrated, and is not a physically motivated boundary.}
    \label{fig:massive quiescent galaxy_AGN_NeutraOut+Hist}
\end{figure*}

More specifically, within our full sample, quenched galaxies residing in the denser environments ($\dn > 1$) have long effective lifetime values: $\teff \sim 0.3-0.5$, meaning that quenching took place a long time (30-50\% of the age of the universe) after the bulk of their formation. This suggests that galaxy quenching in the densest environments proceeds gradually, with star formation declining slowly over an extended period of time. Importantly, this environmental trend for quiescent galaxies is mass-independent, since no significant correlation is found between stellar mass and the environment in which massive quiescent galaxies reside, as discussed in Sect.~\ref{Sec:COSMOS-WEB}. Moreover, in the densest environments quiescent galaxies represent a higher fraction of the population of massive galaxies. This is illustrated in the upper panel of Fig.~\ref{fig:t90t50_D}, showing the massive quiescent galaxy fraction in the Blue Jay sample (which is representative of the general population in terms of colors and star formation rates, unlike the DJA sample). Using the Blue Jay sample, whose selection is unbiased with respect to quiescence, the fraction of massive quiescent galaxies rises from $\approx10\%$ in underdense and average-density regions to $\approx40$--$50\%$ in the densest environments.

In contrast, the star-forming population shows no clear dependence on \dn, regardless of stellar mass. However, galaxies with \ms $ <10^{10}\,$\msun are somewhat more concentrated towards very low values of \teff, suggesting extremely rapid mass assembly. For star-forming galaxies, which are by definition not quenched, \tq\ and \teff\ are less straightforward to interpret than for the quiescent sample: a short \teff\ indicates a system that is rapidly growing, whereas a large \teff\ indicates a star-formation history extended in time. It is therefore interesting that star-forming systems in high-density environments tend to cluster towards relatively high \teff. Such values imply that recent star formation has not played a major role in building up their stellar mass, which suggests early formation followed by a gradual decline in star-formation activity. This is broadly consistent with the behaviour of massive cluster members in the local Universe \citep{2005Thomas}.

\subsection{AGN-driven outflows}\label{quenching mechanisms}

To investigate the physical drivers of quenching in massive systems, we examine several key indicators in Fig.~\ref{fig:massive quiescent galaxy_AGN_NeutraOut+Hist}. We now focus only on the Blue Jay sample, which has robust spectral line measurements from medium-resolution spectroscopy. The left panel shows the massive Blue Jay galaxies from Fig.~\ref{fig:t90t50_D} --- both star-forming and quenched --- with symbols identifying the systems hosting AGN and/or neutral gas outflows. We classify galaxies as AGN using X-ray data and the BPT diagram; and we adopt the neutral outflow measurements based on the Na~D absorption lines provided by \citet{2024Davies}.

Neutral outflows are common among both star-forming and quiescent galaxies. Among the quiescent galaxies with a detected neutral outflow, more than half also show signatures of AGN activity. This points to a likely connection between AGN-driven neutral outflows and galaxy quenching, as has been discussed by \citet{2024Davies} and \citet{2025Bugiani}. We also note that the outflow incidence is substantially higher in quiescent galaxies experiencing rapid quenching (i.e., with short effective timescales \teff), which is consistent with the results of \citet{2024Park}. 

Now, for the first time, we are also able to investigate the relation between AGN-driven outflows and the cosmic environment. Among quiescent galaxies, AGN activity and neutral outflows are found almost exclusively in systems living in low-density environments, with only one AGN host at higher densities. To quantify this, we split the sample at \dn$=0.5$, a value chosen purely to delineate the region where the AGN are concentrated rather than a physically motivated boundary. The right panel of Fig.~\ref{fig:massive quiescent galaxy_AGN_NeutraOut+Hist} shows the number of massive quiescent galaxies hosting AGN and neutral outflows on either side of this split. In the low-density regime (\dn$<0.5$), seven of the eight massive quiescent galaxies show prominent AGN activity and five host strong neutral outflows. In the higher-density regime (\dn$\geq0.5$), by contrast, only one galaxy hosts an AGN and none shows any detectable sign of a neutral outflow.

Our observations thus suggest a picture in which quiescent galaxies in overdense regions quenched over long timescales and are currently devoid of AGN activity and gas outflows, while quiescent galaxies in low-density regions experienced rapid quenching and are currently exhibiting signs of AGN feedback in the form of neutral gas outflows.

\begin{figure*}
\centering
    \includegraphics[width=0.9\linewidth]{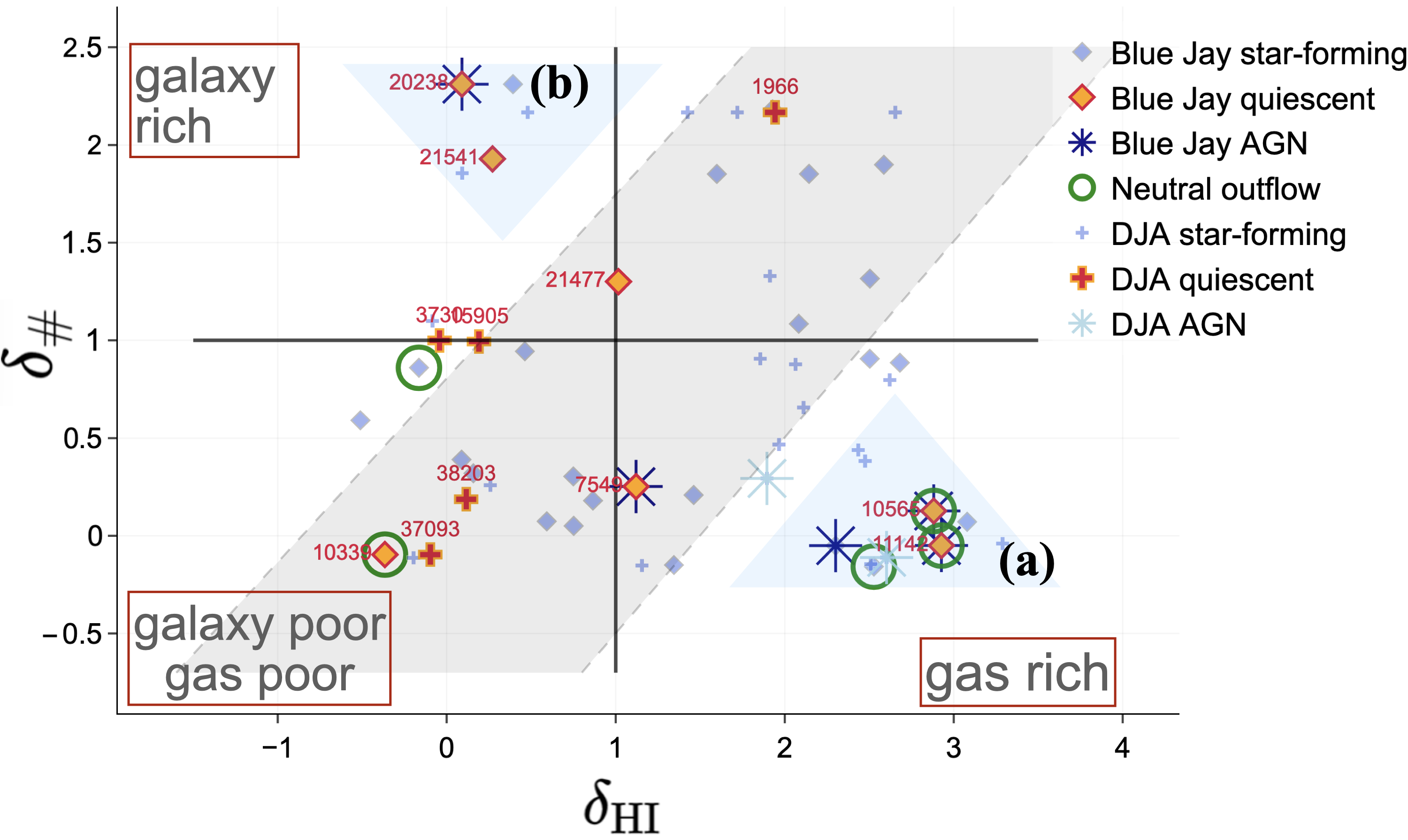}
    \caption{Galaxy number density (\dn, from COSMOS-Web) plotted against neutral-hydrogen density (\dhi, from LATIS Ly$\alpha$ tomography) for the massive galaxies in our sample at $2.2 < z < 2.8$. Roughly 65\% of the sample follows the correlated band between the two tracers, shown in gray. Two regions showing the largest discrepancies between \dhi\ and \dn --- labeled (a) and (b) --- are marked with blue triangles for reference. Horizontal and vertical lines indicate the density thresholds used to examine the SFHs above and below these values, as shown in Fig.~\ref{fig:SFHs}. }
    \label{fig:dh_dn}
\end{figure*}

\begin{figure}
\centering
    \includegraphics[width=0.89\linewidth]{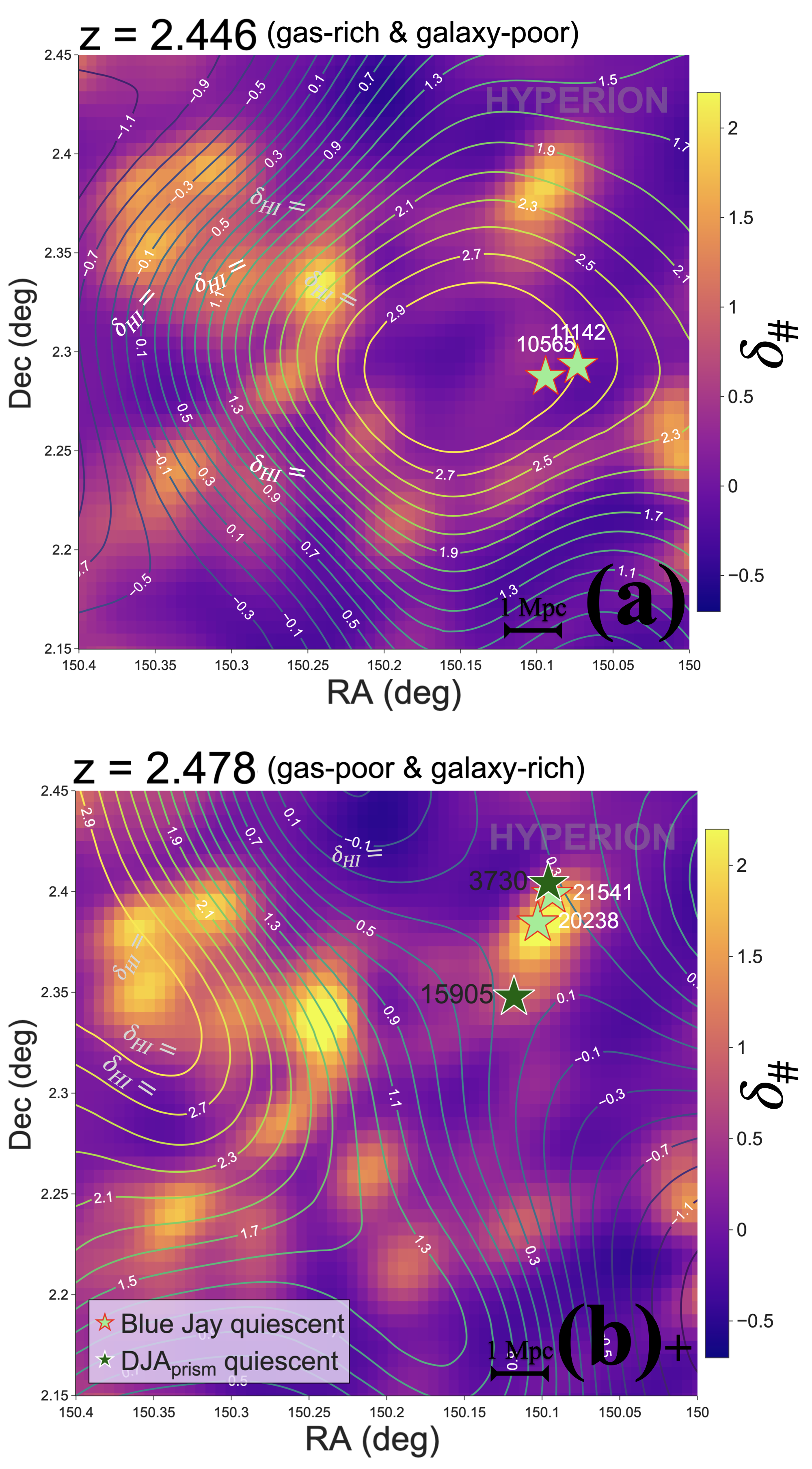}
    \caption{Two redshift slices of the galaxy density map from COSMOS-Web \citep{2026Hatamnia}, coinciding with the massive quiescent galaxies residing in regions (a) and (b) of Fig.~\ref{fig:dh_dn}. The LATIS \citep{2020Newman} tomographic map of neutral hydrogen is overplotted as contours, with \dn\ and \dhi\ values computed following Eq.~\ref{eq:dN} \& \ref{eq:DH}, respectively. }
    \label{fig:hyperion_LATIS}
\end{figure}

\begin{figure*}
\sidecaption
    \includegraphics[width=0.7\linewidth]{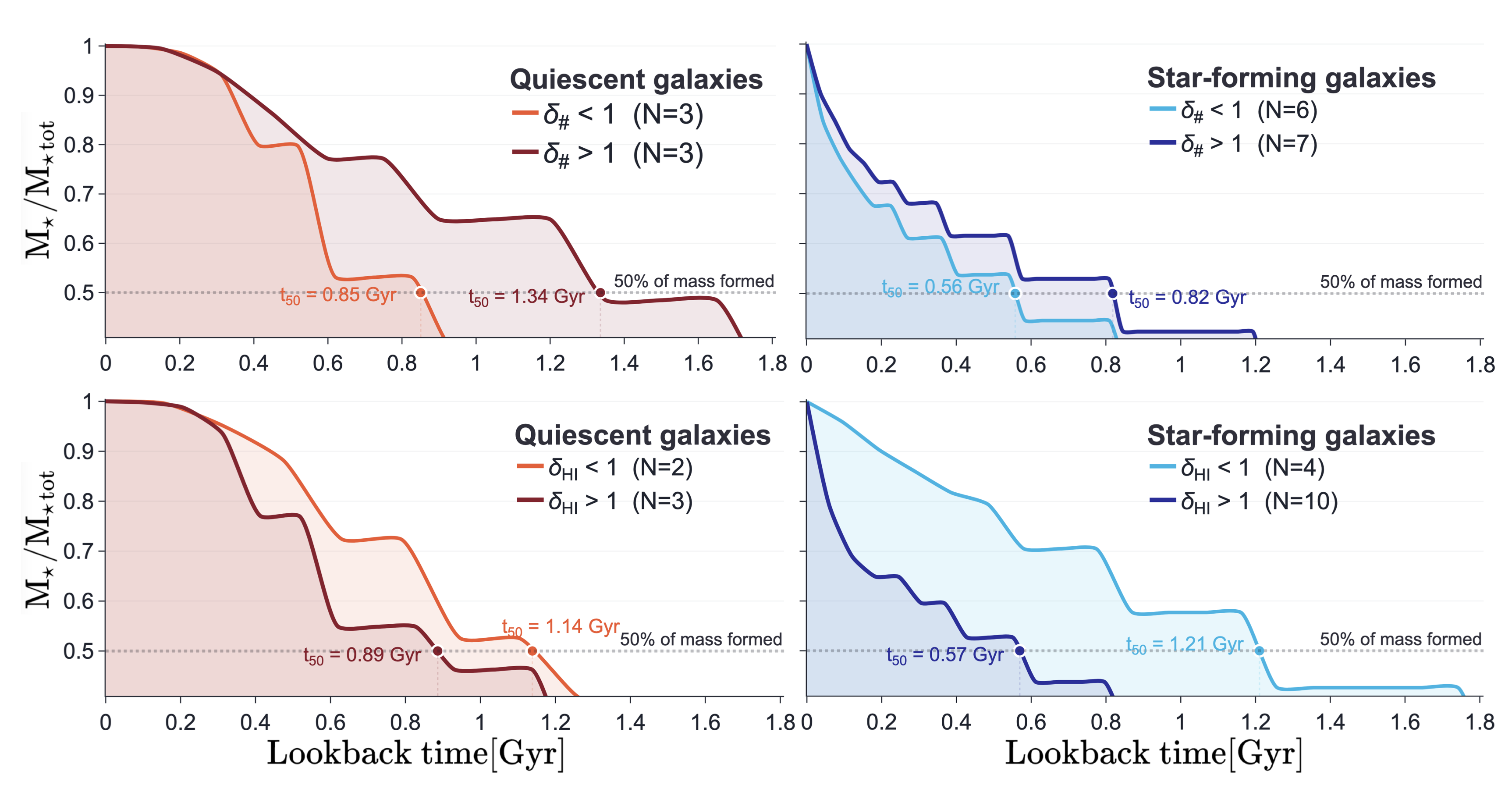}
    \caption{ The stacked (average) mass assembly history of massive galaxies, normalized to their total stellar mass, is shown here as a function of lookback time for the redshift range $2.35$--$2.55$ (roughly $\pm 100$\,cMpc of the Hyperion core). The normalized mass assembly histories of massive quiescent galaxies are shown in the left panels, while the right panels shows the massive star-forming galaxies. The low- and high-density criteria adopted in this plot follow the black solid line in Fig.~\ref{fig:dh_dn}, where \dn$=1$ and \dhi$=1$. The black dashed lines mark \tf, the lookback time at which half of the total $M_\star$ has formed. The number of galaxies in each density regime is given in the legend.} 
    \label{fig:SFHs}
\end{figure*}

\section{Neutral Hydrogen and Number-Density Tracers}\label{Sec:Hyperion Proto-Cluster}\label{Sec:Neutral gas and galaxy number density}

As described in Sect.~\ref{Sec:LATIS}, LATIS performed their HI column density measurements in a limited redshift range of $2.2<z<2.8$. Within this redshift interval, the majority of the observed galaxies lie in the vicinity of the well-known Hyperion protocluster \citep{2018Cucciati}. In particular, the Blue Jay footprint encompasses one of Hyperion's massive cores, whose redshift we estimate to be $z\sim2.47$ based on the spectroscopic redshifts of sample galaxies spatially coincident with this extreme overdensity (i.e., $\dn\gtrsim2$). As a result, 62\% of the massive galaxies in our $2.2<z<2.8$ sample fall within a three-dimensional radius of $\sim$100\,cMpc (i.e., $\sim 25-30 \, $Mpc) of the Hyperion structure. This fraction drops to roughly half among massive star-forming galaxies, yet rises sharply for massive quiescent galaxies, with 11 out of 12 residing within this proximity.

We use \dhi\ (Eq. \ref{eq:DH}) as a proxy for the large-scale neutral hydrogen overdensity. Larger positive values correspond to stronger-than-average Ly$\alpha$ absorption and therefore trace denser H\textsc{I}-rich regions of the intergalactic medium, while negative values indicate weaker absorption and underdense environments. Although the intrinsic nature of the \dhi\ and \dn\ measurements are substantially different, thus preventing a quantitative comparison of their absolute values, we can still learn important information on the relative distribution of galaxies and cool gas by performing a qualitative comparison of these two environmental measurements. We note that, while high values of \dhi\ consistently indicate an overdense region with abundant cool gas, low values are more ambiguous and can arise either from a genuine deficiency of cool gas or from regions where the gas is predominantly in a hotter, ionized phase (e.g., an ionized ICM).

In Fig.~\ref{fig:dh_dn}, we compare these two distinct environment measurements for massive galaxies in our Blue Jay and DJA sample within the redshift 2.2 - 2.8. Considering both star-forming and quenched systems, 65$\%$ of galaxies in our sample show correlated values for \dhi\ and \dn\ as indicated by the grey region in Fig.\ref{fig:dh_dn}; meaning that if  the environment in which they reside is rich in galaxies, then it also shows a larger content of neutral Hhydrogen.

We also show the AGN activity and neutral outflow signatures in Fig.~\ref{fig:dh_dn}. These phenomena preferentially populate the region with low galaxy number density, $\dn<0.5$, as already discussed in Sec.~\ref{Sec:Environment and Formation/Quenching}.

It is thus surprising to see that that every AGN in our sample resides in regions with strong Ly$\alpha$ absorption (i.e., $\dhi >1$) except one (galaxy 20238 in Fig.~\ref{fig:dh_dn}). In fact, AGN activity is detected predominantly in the quadrant of Fig.~\ref{fig:dh_dn} which is galaxy-poor and HI~gas-rich. 

We further investigate the two extreme regions where \dhi\ and \dn\ present the strongest discrepancy. We label these regions (a) and (b), which are highlighted by arbitrary triangular regions in Fig.~\ref{fig:dh_dn}. 
Perhaps the most striking case is region (a), which contains galaxies exhibiting the highest \dhi\ values alongside the lowest \dn\ values. This region also hosts the vast majority of AGNs and more than half of the systems with detectable neutral outflows. We confirm that every massive galaxy in our full sample occupying this region lies within a radius of $\sim$100\,cMpc of the Hyperion protocluster. The two massive quiescent galaxies, in particular, are relatively close ($<1 \,Mpc$), and we show their location in the top panel of Fig.~\ref{fig:hyperion_LATIS}. In the figure, the background heatmap depicts the extended Hyperion structure using the COSMOS-Web density map, and is overlaid with \dhi\ contours tracing the IGM, shown at the specific redshift of the two quenched systems. The galaxies in region (a) appear to be situated within an exceptionally gas-rich bubble (the \dhi = 2.9 contour spans a projected diameter of $\sim$20\,cMpc), which however does not correspond to a peak in the galaxy density map. In fact, these galaxies are adjacent to Hyperion, but outside of its densest parts: the value of \ensuremath{\dn=0} indicates that the local environment has an average density of galaxies.

On the other hand, region (b) is characterized by the highest \dn\ and lowest \dhi. One of those systems show AGN activity, yet, no neutral outflows are detected in this region. In this case, too, the two quiescent galaxies are physically close, and we show their location in the bottom panel of Fig.~\ref{fig:hyperion_LATIS}). The map reveals that these galaxies are situated within one of the largest cores of the Hyperion protocluster; which however presents a clear deficiency of neutral hydrogen. In this regime a genuine deficiency of cool gas cannot be distinguished from a scenario in which gas is present but predominantly ionized, e.g., by a hot intragroup or intracluster medium. Similar maps for all massive quiescent galaxies in the 2.2-2.8 redshift range are available in Fig.~\ref{fig:app_maps}.

\subsection{Galaxy Assembly in the Hyperion Proto-Cluster}\label{Sec:Hyperion Proto-Cluster}

Motivated by the relevance of Hyperion to our results, we now narrow our focus to the immediate surroundings of the protocluster, restricting the analysis of this section to $\Delta z \sim \pm 0.1$ around the main Hyperion core that overlaps with our sample. This constrains the redshift range to $2.35 < z < 2.55$  (roughly within $\pm100$\,cMpc of the Hyperion core). We emphasize that our sample covers only a single massive core within the Hyperion structure \citep{2018Cucciati,2026Sikorski}, and the resulting trends should not be generalized to the structure as a whole, which likely spans a range of masses, gas temperatures, and gas contents across its cores.

In order to study the timescales of galaxy formation, in Fig.~\ref{fig:SFHs} we present the stacked mass assembly histories as a function of lookback time, normalized to the total stellar mass, for massive galaxies within the redshift range $2.35 < z < 2.55$. The normalization enables a direct comparison of the shape of the mass growth curves (i.e., the cumulative star formation histories) across different environments, in a way that is independent of absolute stellar mass. Massive quiescent galaxies and massive star-forming galaxies are shown separately in the left and right panels, respectively, with each population further split into low- and high-density subsamples following the thresholds \dn$=1$ and \dhi$=1$ (marked by the black lines in Fig.~\ref{fig:dh_dn}). Each panel highlights the normalized mass assembly beginning just before $\mathrm{M_\star/M_{\star,total}} = 0.5$, which marks the galaxy's formation time, \tf. The ``staircase'' appearance of the stacked mass assembly histories reflects the time bins of the non-parametric SFH derived with Prospector, as explained in Sec.~\ref{Sec:Data}.

The left panels of Fig.~\ref{fig:SFHs} reveal that massive quiescent galaxies form significantly earlier in high-density environments, with formation times nearly double those of their low-density counterparts. The opposite, though less pronounced, trend emerges in the lower panel: massive quiescent galaxies residing in regions of higher neutral-gas density form a few hundred Myr later than those in gas-deficient regions. This reversal is consistent with the distribution of massive quiescent galaxies already seen in Fig.~\ref{fig:dh_dn}. In both cases, however, the final stages of massive quiescent galaxy evolution appear similar, with quenching well underway (\ms/M$_{\star tot}>0.9$) and the mass assembly rate reduced to a minimum.

In contrast, the right panels of Fig.~\ref{fig:SFHs} reveal a markedly different trend for massive star-forming galaxies. Their formation appears far more sensitive to the cool gas content of their environment, as evidenced by the small number of massive star-forming galaxies found in low-\dhi\ regions and their correspondingly longer \tf\ --- roughly double that of their high-\dhi\ counterparts. It is also notable that, while the mass assembly rate of massive star-forming galaxies is comparable between low- and high-number-density regions, it becomes strikingly elevated in cool-gas-rich environments, where on average a quarter of the total stellar mass has been assembled within just the last 100\,Myr. Taken together, these results underscore the importance of considering multiple environmental tracers.

\section{Discussion}\label{Sec:Discussion}

\subsection{The impact of the environment on the star formation history}

One of our main results is that massive quiescent galaxies that live in environments with a high overdensity of galaxies (\ensuremath{\dn>1}) present a systematically different star formation history compared to their counterparts living in regions with average or low density. This is clearly shown for our full sample in Fig.~\ref{fig:t90t50_D}: in overdense regions massive galaxies quenched a long time (up to half of the age of the universe) after the bulk of their stars were formed. A more direct comparison of the star formation histories can be performed by selecting galaxies observed at the same redshift; this is shown in the top left panel of Fig.~\ref{fig:SFHs} for the subset of galaxies that are located within the Hyperion region. It can be clearly seen that, among massive quiescent galaxies, those living in overdense regions formed more than 0.5~Gyr earlier compared to the others.

There are two important caveats when interpreting the star formation histories inferred from spectro-photometric fits. First, the accuracy of the reconstruction steadily decreases at older ages; this is due to the less rapid spectral evolution of older stars, together with the outshining effect by brighter, younger stars. For this reason we show only the final 60\% of the mass formation in Fig.~\ref{fig:SFHs}, and we do not attempt to constrain the earliest phases of the star formation histories. Second, as quiescent galaxies age they accrete a progressively larger number of stars that formed in satellite systems. The contribution of these accreted stars to the star formation history cannot be disentangled from the truly in-situ star formation. It is thus possible that the slow evolution observed in massive quiescent galaxies living in overdense regions is, at least partially, produced by the accretion of younger stars, rather than to prolonged star formation. Nonetheless, these caveats do not substantially alter our conclusions: in overdense regions, massive quiescent galaxies form earlier. Our findings are also consistent with observations at somewhat lower redshift. \citet{2020Webb} studied massive quiescent galaxies at $1 < z < 1.5$ and found that those in clusters are older than their field counterparts at fixed stellar mass, with a mean mass-weighted age difference of $\sim0.31$\,Gyr, corresponding to earlier formation in overdense environments.

Moreover, we checked that this trend does not simply reflect a mass trend (i.e., overdensities contain more massive galaxies, and more massive galaxies form earlier); as shown in Fig.~\ref{fig:mass delta}, the stellar mass of massive quiescent galaxies in our sample does not correlate with the local overdensity as measured by $\dn$. These findings are consistent with those of \citet{2024Taamoli} and \citet{2026Hatamnia}, who found that the stellar masses of quiescent galaxies depend, if any, only weakly on local galaxy density. Using the COSMOS2020 \citep{2022Weaver} and COSMOS-Web \citep{2023Casey} catalogues, they showed that at $2 \lesssim z \lesssim 3$, environment has little effect on the stellar mass distribution of the quiescent population, whereas more massive star-forming galaxies are preferentially found in denser regions.

\subsection{Galaxies versus gas as environment tracers}
\label{sec:disc_tracers}

Roughly two thirds of our sample in the redshift range $2.2 < z < 2.8$ follows a correlation between \dhi and \dn (gray band in Fig.~\ref{fig:dh_dn}), as expected if both galaxies and cool gas trace the underlying matter distribution. The remaining third does not follow this correlation, and the two extreme regimes, gas-rich but galaxy-poor, and galaxy-rich but gas-poor, are the most informative part of the comparison. Before interpreting them physically, we consider whether they could arise from the measurements themselves \citep[as suggested in a number of studies such as ][]{2015Stark,2018Krolewski,2022Newman}.

The maps of the two types of measurements, i.e., \dn\ and \dhi, have spatial scales that are comparable on transverse scales ($\sim1.7\,$cMpc and $\sim4\,$cMpc respectively); and we verified that smoothing the \dn\ map to the \dhi\ resolution leaves Fig.~\ref{fig:dh_dn} essentially unchanged, with the same galaxies populating regions (a) and (b). However, the resolution differs along the line of sight, where the LATIS tomographic reconstruction, which is based on spectroscopic data, has a higher resolution ($\sim 17\, $cMpc), as opposed to the COSMOS-Web \dn\ map ($\sim 50\, $cMpc) which is limited by the uncertainty on the photometric redshifts. 
A structure more compact than the line-of-sight resolution is therefore diluted in \dn\ while remaining resolved in \dhi, which would mimic the signature of region (a) in Fig.~\ref{fig:dh_dn} without any decoupling between gas and galaxies.                                                                                                                                                                                                  The two density tracers, \dn\ and \dhi, have comparable resolution on transverse scales, and we verified that smoothing the \dn\ map to the \dhi\ resolution leaves Fig.~\ref{fig:dh_dn} essentially unchanged, with the same galaxies populating regions (a) and (b). Along the line of sight, however, the LATIS reconstruction is nearly isotropic ($\sigma\approx7\,$cMpc; \citealt{2020Newman}), whereas the COSMOS-Web \dn\ map is computed in slices $35\,h^{-1}\,$cMpc ($\approx50\,$cMpc) thick \citep{2024Taamoli}. One could therefore suspect that the galaxies in region (a) lie in compact overdensities that are diluted into apparent underdensities in \dn, while remaining resolved in \dhi. This effect is partially mitigated by the wKDE adopted for COSMOS-Web, which down-weights galaxies with poorly constrained redshifts rather than spreading them uniformly \citep{2016Darvish,2024Taamoli}.

To directly assess the impact of photometric-redshift uncertainties, we turn to the Hyperion density reconstruction from the Charting Cluster Construction with VUDS and ORELSE (C3VO) survey \citep{2022Lemaux, 2025Hung}, which is partially anchored by spectroscopic redshifts. We show a snapshot of the galaxy number density reconstruction in Fig.~\ref{fig:3D}; the full animation is available online\footnote{\url{https://www.youtube.com/watch?v=TiQhA-0q14A}}. We find good qualitative agreement with the COSMOS-Web map at $2.3<z<2.7$, around the Hyperion protocluster.
In particular, the Hyperion density field reconstructed by \citet{2026Sikorski} with a Voronoi Monte Carlo tessellation of a combination of spectroscopic and photometric redshifts shows qualitatively the same structures as the density maps used in this work, providing an independent check of our results. Matched against their seven $\mathrm{5\sigma_{\delta}}$ peaks, our region (b) in Figs.~\ref{fig:dh_dn} \& \ref{fig:app_maps} fall on the most massive peak, ``Theia'' ($z=2.468$), while the galaxies in region (a) lie several arcminutes from the nearest peak, ``Helios'' ($z=2.444$), at essentially the same redshift but outside its contour, in the filamentary regions connecting the peaks. Thus, the reconstruction of \citet{2026Sikorski}, which is partially anchored by spectroscopic redshifts and therefore less affected by photometric-redshift uncertainties, supports our results by placing region (a) in a \emph{truly} low-density environment adjacent to the massive cores, and region (b) in a high-density environment. We further test the effect of spectroscopic redshifts in Appendix~\ref{app:LBG}, where we use the LATIS catalog of LBGs to derive the galaxy number density, and show that the gas-rich, galaxy-poor environments are qualitatively unchanged.

\begin{figure}
    \centering
    \includegraphics[width=\linewidth]{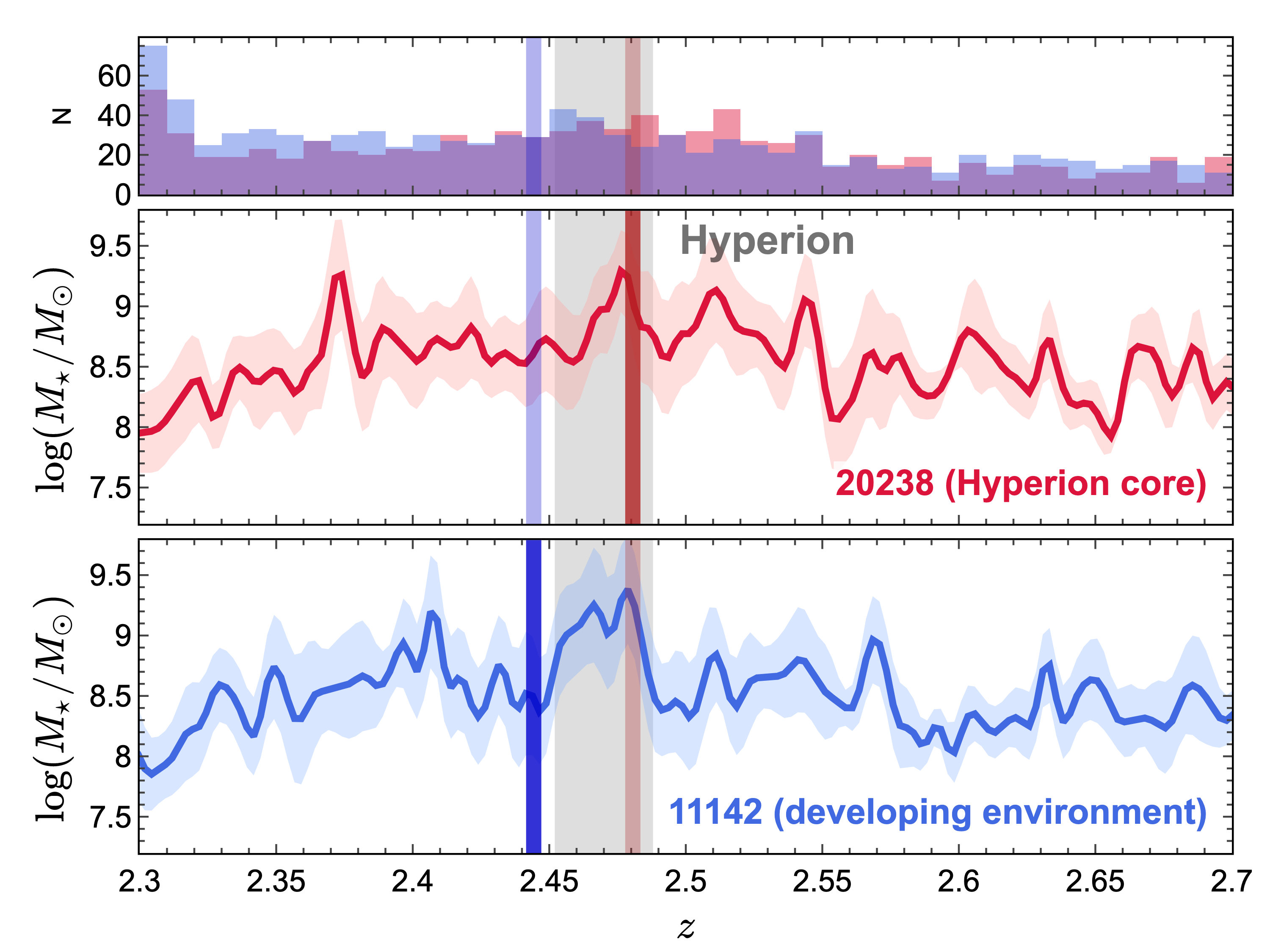}
    \caption{COSMOS-Web \citep{2023Casey} stellar-mass distribution along the line of sight around two of our massive quiescent galaxies: 20238 (red, resides in a Hyperion core) and 11142 (blue, resides in a developing environment). For each target, galaxies are selected within a cylinder of 2\,Mpc radius centered on its sky position, and are shown as a function of redshift. \textit{Bottom:} median stellar mass of galaxies in each slice (solid lines), with shaded regions showing the $1\sigma$ scatter. The vertical lines mark the redshifts of 20238 and 11142, and the gray band indicates the redshift range of the Hyperion protocluster. \textit{Top:} number of galaxies per redshift slice.}
    \label{fig:mass_dist}
\end{figure}

Gas-rich, galaxy-poor environments such as region (a) have been proposed as candidate UV-dim protoclusters \citep[e.g.,][]{2022Newman, 2025Newman}, in which most member galaxies are too faint in the rest-frame UV to be detected by UV-selected surveys, for instance because they are dust-obscured or dominated by older stellar populations. In this picture, strong \ion{H}{i} absorption on the scale of a massive halo, without a corresponding galaxy overdensity, reflects the limitations of UV-based galaxy selection rather than a genuine absence of galaxies. In our case, the sensitive near-infrared data from JWST NIRCam makes this concern less relevant: at $z\sim2.5$, the COSMOS-Web catalog includes galaxies that are faint in the UV, down to a stellar-mass completeness of $\sim10^{8}\,\msun$ \citep{2023Casey,2026Hatamnia}.
We thus conclude that region (a) truly lacks an overdensity of galaxies; however we note that it is $<60$ cMpc to one of the largest known protoclusters at these redshifts, and is therefore not in an average cosmic environment (on large scales). This is illustrated in Fig.~\ref{fig:mass_dist}, which shows the average stellar-mass distribution of COSMOS-Web galaxies within a $2$\,Mpc-radius cylinder around two massive quiescent galaxies: 20238, in region (b), and 11142, in region (a). The environment at the redshift of 11142 (lower panel) has an average stellar mass roughly an order of magnitude lower than the overdense environment of 20238 (middle panel), which overlaps with an extended Hyperion core.

On the other hand, the low \dhi\ measured in region (b) has two possible explanations: it may signal the absence of gas, or it is possible that the gas in that region is in a hot, ionized phase invisible to the Ly$\alpha$ forest. The coincidence of region (b) with one of the Hyperion collapsed cores makes the ionized interpretation plausible: these galaxies are not necessarily gas-poor in an absolute sense, but reside in regions where the neutral-gas tracer no longer traces the total gas content.

\subsection{AGN, outflows, and the cool gas supply}
\label{sec:AGN, outflows, and the cool gas supply}

In our sample, galaxies hosting neutral outflows or AGN activity are preferentially found in the region of the parameter space with low \dn\ and high \dhi. This is a striking result: nuclear activity and outflowing neutral gas are found not in the galaxy-rich cores of Hyperion, but in the gas rich, galaxy-poor regions surrounding them. 
It is particularly suggestive to note that one of the two quiescent galaxies found in the gas-rich, galaxy-poor region (a) is 11142. This object was studied in detail in \citet{2024Belli} as it represents one of the clearest cases of galaxy quenching by AGN ejective feedback. The galaxy appears to be in the declining phase of a particularly rapid quenching, and hosts a multi-phase outflow which is very effective at removing the cold neutral ISM. Remarkably, the other galaxy found in region (a) is 10565, which shares many similarities with 11142, and hosts the neutral outflow with the highest mass-loading factor of the entire Blue Jay sample \citep{2024Davies}. 

If this correlation is not a coincidence, it requires a plausible explanation for the connection between kpc-scale outflows of neutral gas and the presence of cool gas on Mpc scales.
One plausible interpretation is that AGN triggering in this regime is connected to cool-gas accretion along filamentary structures rather than to galaxy-galaxy interactions or mergers, which are typically more efficient in higher-density group environments \citep[see ][]{2026Giddings}. The cool streams that feed such regions \citep{2005Keres, 2009Dekel} supply both the fuel for black hole growth and the neutral material that is subsequently entrained in the outflow, which would naturally account for the coincidence of the two signatures in these galaxies. Alternatively, outflows may simply be easier to launch and to detect in less crowded environments, where galactic winds encounter a lower-density CGM and IGM and less confinement from neighbouring structures. In any case, the connection with the large-scale environment would be crucial for understanding both the rapid formation and the subsequent quenching of massive galaxies at cosmic noon and beyond.

We caution that these trends rest on a limited number of AGN ($N=6$) drawn from a single massive core of a structure extending over tens of comoving megaparsecs, and we therefore present them as suggestive rather than conclusive. A handful of X-ray selected AGN have been reported elsewhere in the extended Hyperion structure \citep{2022Huang}, in regions our field of view does not reach. We note, however, that the number of AGN detected across the full structure is comparable to the number we recover in the vicinity of a single core. The concentration of AGN we report is therefore unlikely to be an artifact of our restricted coverage.

\subsection{Developing environments}
\label{Developing environments}

Taken together, the preceding sections describe a class of cosmic structures, such as region (a), that are not well captured by either end of the conventional environmental axis. These regions are rich in neutral hydrogen (cool gas) but poor in galaxies; they host the majority of the AGN and neutral outflows in our sample; their massive star-forming galaxies are assembling a quarter of their stellar mass within the last 100\,Myr; and they lie within $\sim100$\,Mpc of a collapsing protocluster core without being part of it. They are neither mature protocluster cores, in which the gas has been processed and the massive population is already quenched, nor low-\dn, low-\dhi field environments, in which neither the fuel nor the structure is present.
 
We introduce the term \emph{developing environment} to describe this phase of cosmic structure formation: a regime in which massive haloes are actively assembling new massive galaxies, still fed by the cool gas of the surrounding web, with black hole growth and feedback underway but the collapse of the structure incomplete. In this framing, the sequence from developing environment to mature core is not a sequence of increasing galaxy density alone, but of the progressive conversion of the neutral reservoir, first into stars, then into outflows, and finally into a hot ionized phase that renders it invisible to Ly$\alpha$ tomography. Region (a) and region (b) of Fig.~\ref{fig:dh_dn} would then represent two snapshots of the same process, separated by several hundred Myr.

A comparable case is the ``Red Potato'' at $z=3.25$ \citep{2026Wang}, a massive quiescent galaxy embedded in a gas-rich cosmic-web node. In that system, the suppression of star formation has been attributed to a luminous X-ray AGN $\sim$48\,kpc away, whose jet may be driving turbulence in the circumgalactic medium and preventing cool gas from accumulating onto the galaxy \citep[also known as \emph{Dragon effect},][]{2026khoram}. In contrast, none of the quiescent galaxies in region~(a) has a bright AGN companion at comparable separation, so a neighbouring jet cannot be invoked to explain their lack of star formation despite their neutral-gas-rich large-scale environment, while,  majority of our quiescent systems are powerful AGN themselves, in these regions. We note that the two measurements trace different scales: \citet{2026Wang} probe the CGM on tens of kpc scales, while \dhi\ traces the IGM on Mpc scales, so the comparison is qualitative.

These observations can be placed in the context of recent theoretical work, although the models do not yet agree on where early massive quiescent galaxies should reside. In the Magneticum Pathfinder simulations, \citet{2025Kimmig} find that massive quenched galaxies at $z \sim 3.4$ preferentially occupy regions that are underdense relative to galaxies of similar stellar mass. \citet{2025Kimmig} attribute this to continued cool-gas accretion in dense environments. Cool gas can keep streaming in along the cosmic web because a stable hot halo has not yet formed to heat or block it. After an AGN outburst temporarily suppresses star formation, this fresh gas can therefore refuel the galaxy and allow star formation to resume. In the SIMBA-C simulation, \citet{2025Szpila} find that early quenched galaxies are not atypical in environment or halo mass at the time of quenching, but tend to have grown their black holes rapidly in dense surroundings before moving into moderately dense regions, where jet-mode feedback can quench them effectively. The GAEA semi-analytic model \citep{2026DeLucia} instead places massive quiescent galaxies at $3 \lesssim z \lesssim 5$ across the full range of environments, from void-like regions to dense filament intersections, and cautions against assuming that all such systems trace the progenitors of present-day massive clusters. Our developing environments fit most naturally within the picture emerging from Magneticum and SIMBA-C: In dense, gas-rich regions, the halo gas may not yet have been heated to a hot, virialized state. Cold gas can then continue to flow onto the galaxy, sustaining or even restarting star formation despite AGN feedback. In such environments, quenching is expected to become lasting only once the neutral gas reservoir has been consumed by star formation, expelled by outflows, or heated. Until then, continued accretion of cool gas from the surrounding may still rejuvenate star-formation.

\section{Summary}
\label{sec:summary}
 
We have combined JWST/NIRSpec spectroscopy of 141 galaxies from the Blue Jay survey with 57 massive galaxies with DJA prism spectroscopy in the COSMOS field, and compared their inferred star formation histories, AGN activity, and neutral outflow signatures against two independent tracers of environment: the COSMOS-Web galaxy number-density field \citep[][ covering the whole redshift range of the sample, 1.7<z<3.5]{2026Hatamnia} and the LATIS Ly$\alpha$ forest tomographic map \citep[][ limited to the radshift range of 2.2-2.8]{2025Newman}. Our main findings are:
 
\begin{itemize}

   \item The fraction of massive quenched galaxies rises from $\sim10\%$ in underdense and average-density regions to $40$--$50\%$ in the densest environments probed. This trend is independent of stellar mass.
 
    \item The normalised interval between formation and quenching increases with galaxy number density. Consistently, stacked mass assembly histories indicate that massive quiescent galaxies in dense regions formed nearly twice as early as their counterparts in low-density regions. These results suggest that the environmental dependence of quenching at cosmic noon is driven primarily by differences in formation time, rather than by differences in the time elapsed between formation and quenching.
 
   \item Formation epoch and mass assembly of massive star-forming galaxies show no dependence on galaxy number density but a strong dependence on cool gas density, assembling on average a quarter of their stellar mass within the last 100\,Myr in the most gas-rich environments.
 
    \item Signatures of AGN activity are found almost exclusively in galaxies with $\delta_{\#}<0.5$ and $\delta_{\rm HI}>1$, with only one exception, while neutral outflow signatures are restricted to galaxies with $\delta_{\#}<1$. Only massive quiescent galaxies exhibit both signatures simultaneously, suggesting a link between nuclear activity, neutral gas ejection, and the quenching of star formation.
 
    \item Approximately $65\%$ of massive galaxies follow a broad correlation between the two, galaxy count and cool gas, tracers while the remaining galaxies deviate from this relation. These outliers are spatially coherent and show systematic differences in their assembly histories, suggesting that the deviations reflect genuine physical differences in galaxy evolution rather than instrumental effects.
 
   \item We identify a regime of gas-rich, galaxy-poor regions adjacent to a Hyperion core in which massive haloes are actively assembling massive galaxies, which we term \emph{developing environments}.
\end{itemize}
 
These results underscore the importance of considering multiple environmental tracers: galaxy counts alone would classify regions (a) and (b) of Fig.~\ref{fig:dh_dn} as, respectively, unremarkable and dense, and would miss the distinction that our analysis suggests is the more physically meaningful one. The joint analysis of galaxy density and cool gas distribution in this work revealed distinctions that crucial for understanding how environment regulates gas accretion, star formation, and ultimately galaxy quenching.


\begin{acknowledgements}

S.B. acknowledges useful discussions with Greg Rudnick and Sebastiano Cantalupo.
AK, SB, LB, CL, MS, GM, and SS are supported by the the
ERC Starting Grant “Red Cardinal”, GA 101076080.
This work is based on observations made with the
NASA/ESA/CSA James Webb Space Telescope. The
data were obtained from the Mikulski Archive for Space
Telescopes at the Space Telescope Science Institute,
which is operated by the Association of Universities for
Research in Astronomy, Inc., under NASA contract NAS
5-03127 for JWST. These observations are associated
with program GO 1817. This work also makes use of
observations taken by the 3D-HST Treasury Program
(GO 12328) with the NASA/ESA HST, which is operated by the Association of Universities for Research in
Astronomy, Inc., under NASA contract NAS5-26555. A.B.N. and H.H. acknowledge the support from Grant 63667 from the John Templeton Foundation. Any opinions, findings, and conclusions or recommendations expressed in this material are those of the authors and do not necessarily reflect the views of the John Templeton Foundation.

\end{acknowledgements}

\clearpage

\bibliography{ref.bib}{} 
\bibliographystyle{aasjournal}

\begin{appendix}
\nolinenumbers

\section{Additional environment maps}
\label{app:maps}

\begin{table}
\centering
\caption{Properties of the Blue Jay quiescent sample.}
\label{tab:BJ_Tab}
\rowcolors{2}{gray!12}{white}
\renewcommand{\arraystretch}{1.5}
\begin{tabular}{lccccccc}
\toprule
ID & z & log(\ms/\msun) & \dn  & \dhi & teff & AGN & Neutral outflow \\
\midrule
8013 & 1.69 & $10.67 \pm 0.02$ & 0.45 & -- & $0.29^{+0.12}_{-0.09}$ & \checkmark & \checkmark \\
10592 & 1.80 & $11.26 \pm 0.02$ & 1.05 & -- & $0.07^{+0.03}_{-0.02}$ & -- & -- \\
10128 & 1.85 & $11.33 \pm 0.03$ & 0.72 & -- & $0.18^{+0.08}_{-0.07}$ & -- & -- \\
19572 & 1.87 & $10.90 \pm 0.03$ & 0.03 & -- & $0.27^{+0.14}_{-0.10}$ & \checkmark & -- \\
8469 & 1.87 & $10.66 \pm 0.02$ & 0.89 & -- & $0.35^{+0.09}_{-0.08}$ & -- & -- \\
16419 & 1.93 & $11.42 \pm 0.02$ & 0.04 & -- & $0.11^{+0.14}_{-0.04}$ & \checkmark & -- \\
11494 & 2.09 & $11.82 \pm 0.02$ & 0.39 & -- & $0.32^{+0.08}_{-0.07}$ & \checkmark & \checkmark \\
9395 & 2.13 & $10.80 \pm 0.02$ & 0.91 & -- & $0.22^{+0.09}_{-0.07}$ & -- & -- \\
10339 & 2.36 & $10.54 \pm 0.03$ & -0.10 & -0.37 & $0.13^{+0.07}_{-0.03}$ & -- & \checkmark \\
10565 & 2.44 & $10.84 \pm 0.03$ & 0.13 & 2.88 & $0.23^{+0.13}_{-0.09}$ & \checkmark & \checkmark \\
11142 & 2.44 & $10.95 \pm 0.03$ & -0.05 & 2.93 & $0.07^{+0.02}_{-0.01}$ & \checkmark & \checkmark \\
21477 & 2.47 & $10.83 \pm 0.04$ & 1.30 & 1.02 & $0.15^{+0.18}_{-0.07}$ & -- & -- \\
21541 & 2.48 & $11.05 \pm 0.02$ & 1.93 & 0.27 & $0.31^{+0.10}_{-0.06}$ & -- & -- \\
20238 & 2.49 & $11.10 \pm 0.10$ & 2.31 & 0.09 & $0.31^{+0.14}_{-0.16}$ & \checkmark & -- \\
7549 & 2.62 & $10.86 \pm 0.02$ & 0.25 & 1.12 & $0.10^{+0.20}_{-0.05}$ & \checkmark & -- \\
\bottomrule
\end{tabular}
\end{table}
\begin{table}
\centering
\caption{Properties of the DJA quiescent sample.}
\label{tab:mqg}
\rowcolors{2}{gray!12}{white}
\renewcommand{\arraystretch}{1.5}
\begin{tabular}{lccccccc}
\toprule
ID & z & log(\ms/\msun) & \dn & \dhi & teff & AGN & Neutral outflow \\
\midrule
38203 & 2.24 & $11.26^{+0.05}_{-0.01}$ & 0.19 & 0.12 & $0.29^{+0.11}_{-0.07}$ & \textit{N/A} & \textit{N/A} \\
37093 & 2.37 & $10.86^{+0.12}_{-0.04}$ & -0.10 & -0.10 & $0.13^{+0.34}_{-0.08}$ & \textit{N/A} & \textit{N/A} \\
1966 & 2.47 & $11.40^{+0.09}_{-0.05}$ & 2.17 & 1.94 & $0.20^{+0.06}_{-0.02}$ & \textit{N/A} & \textit{N/A} \\
15905 & 2.48 & $11.03 \pm 0.01$ & 0.99 & 0.19 & $0.30^{+0.06}_{-0.03}$ & \textit{N/A} & \textit{N/A} \\
3730 & 2.48 & $10.91 \pm 0.01$ & 1.00 & -0.04 & $0.39^{+0.02}_{-0.03}$ & \textit{N/A} & \textit{N/A} \\
\bottomrule
\end{tabular}
\end{table}

\begin{figure*}
\centering
    \includegraphics[width=\linewidth]{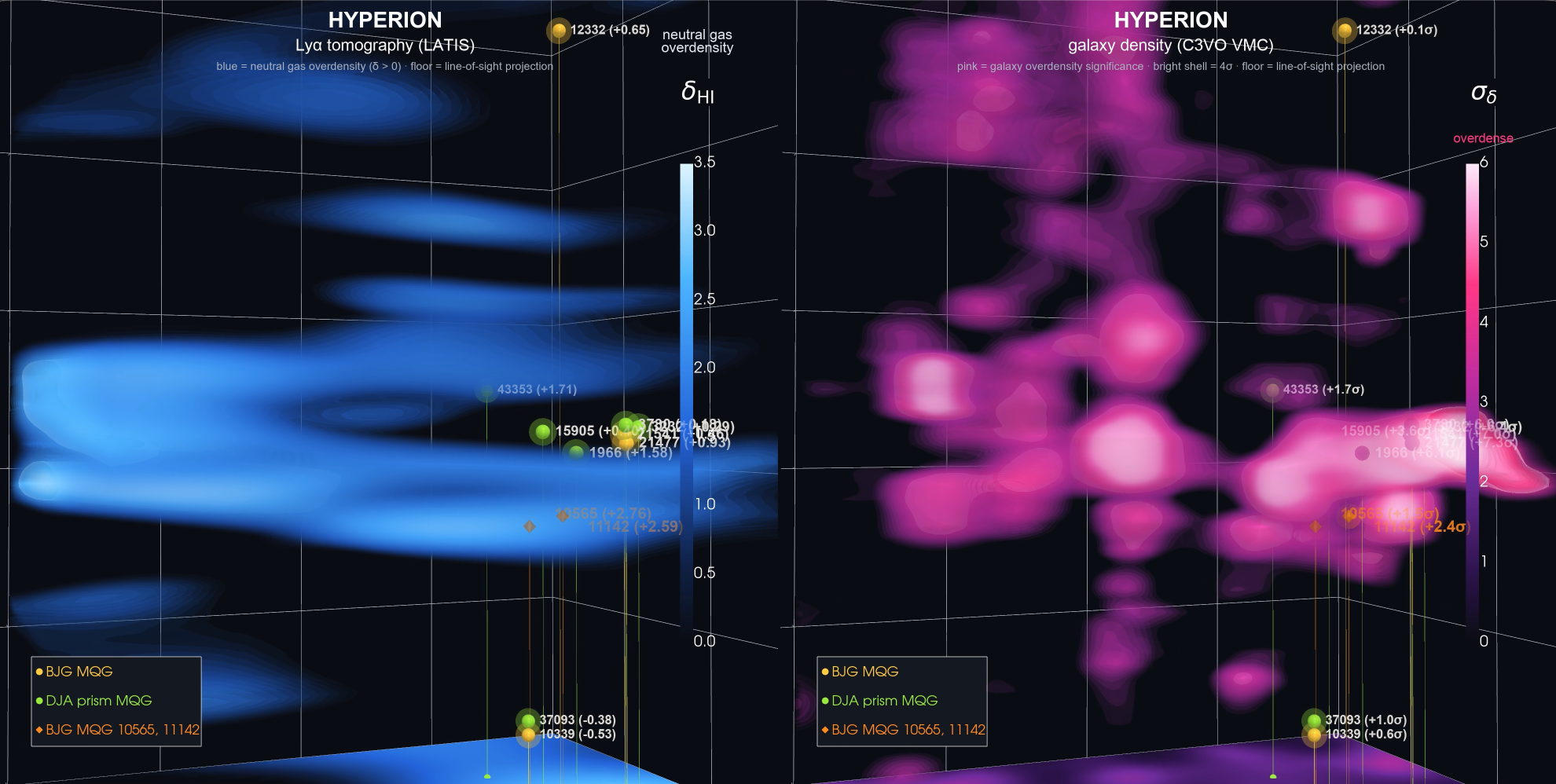}
    \caption{A snapshots of the LATIS cool-gas and C3VO galaxy density distributions, with massive quiescent galaxies located near the Hyperion protocluster. Blue traces in the left panel illustrate excess cool gas (\dhi > 0), while the right panel depics galaxy densities down to the 1.5$\sigma$ contour. The full video is publicly available at the link \url{https://www.youtube.com/watch?v=TiQhA-0q14A}.}
    \label{fig:3D}
\end{figure*}

\begin{figure}
\sidecaption
    \includegraphics[width=1.\linewidth]{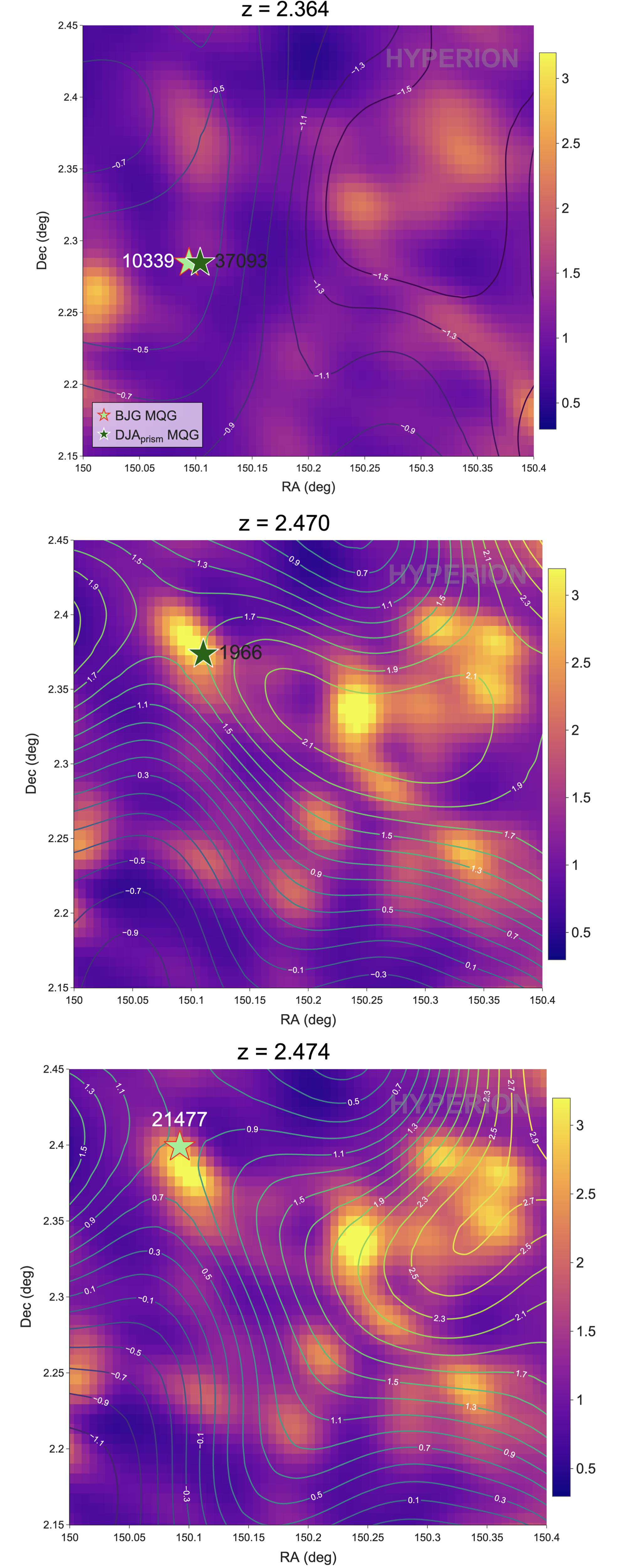}
    \caption{Three redshift slices of the galaxy density map \citep{2026Hatamnia}, each centered on the redshift of an massive quiescent galaxy, as indicated at the top of each panel. Contours show the LATIS \citep{2020Newman} tomographic map of neutral hydrogen, with \dn\ and \dhi\ values computed from Eqs.~\ref{eq:dN} and \ref{eq:DH}, respectively.}
    \label{fig:app_maps}
\end{figure}

\clearpage 

\clearpage

\section{Comparison to Lyman-break galaxy counts}
\label{app:LBG}

One limitation of the COSMOS-Web catalog is that it is based on photometric redshifts, which have relatively low precision. As a result, the \dn\ signal can be smoothed along the line-of-sight direction. It is thus useful to test whether our results change when using a catalog with spectroscopic redshifts. We do this by adopting the LATIS catalog of LBGs \citep{2025Newman}, for which the redshifts were spectroscopically measured. This catalog offers a superior resolution along the line-of-sight, but is limited to a small subset of galaxies in any given region.

We derive $\delta_\mathrm{LBG}$, a measure of overdensity based on LBG galaxy counts, by adopting a 3D isotropic Gaussian kernel with $\sigma = 4 $~cMpc/h, following \citep{2025Newman}. In 3D, a sphere with radius 1$\sigma$ encloses only 20\% of the kernel weight, while a radius of 1.87$\sigma$, corresponding to 7.5~cMpc/h, encloses 68\% of the kernel weight. This means that the resulting $\delta_\mathrm{LBG}$ map is smoothed on large spatial scales, which is a requirement driven by the low surface density of LBGs. On the contrary, the dense sampling offered by the large COSMOS-Web catalog, together with the adaptive kernel size adopted when deriving \dn, enable a higher spatial resolution. The LBG and COSMOS-Web catalogs thus suffer from complementary limitations.

We compare $\delta_\mathrm{LBG}$ and \dhi\ for our sample in Fig.~\ref{fig:3D}. This is analogous to Fig.~\ref{fig:dh_dn}, with the difference that the galaxy number density is now estimated from the LBG catalog instead of the COSMOS-Web one. Most of the sample follows a trend between  $\delta_\mathrm{LBG}$ and \dhi, which we highlight with a gray band. Interestingly, the two quiescent galaxies 10565 and 11142 are substantially below the trend, thus confirming the ``gas-rich, galaxy-poor'' nature of their environment. One difference with Fig.~\ref{fig:dh_dn}, however, is that the galaxy overdensities based on LBGs are much higher than those based on COSMOS-Web, extending up to $\delta_\mathrm{LBG}\sim14$, reflecting mostly the different spatial resolution of the two maps rather than the nature of the tracer, which is biased towards star-forming galaxies in a specific mass range. Even the two galaxies 10565 and 11142 present a substantial overdensity of $\delta_\mathrm{LBG}\sim 6$; however, they still live in a galaxy-poor environment when compared to other galaxies at the same \dhi, such as 1966. Moreover, all galaxies with detected AGN activity or neutral outflows other than 11142 and 10565 are confined to $\delta_\mathrm{LBG}\lesssim4$. Thus, the gas-rich, galaxy-poor aspect remains confirmed in a relative sense, and our main conclusions are not affected.

Finally, we note that galaxy 20238, which lives in one of the densest cores of Hyperion, has an extremely low $\delta_\mathrm{LBG}$ despite being in a galaxy-rich environment according to \dn. This could be attributed to systematic differences in galaxy properties: it is possible that in the Hyperion core galaxies are more mature, thus more massive and less star-forming, and therefore the number of LBGs detected in this environment is drastically lower.

\begin{figure*}
\centering
    \includegraphics[width=0.8\linewidth]{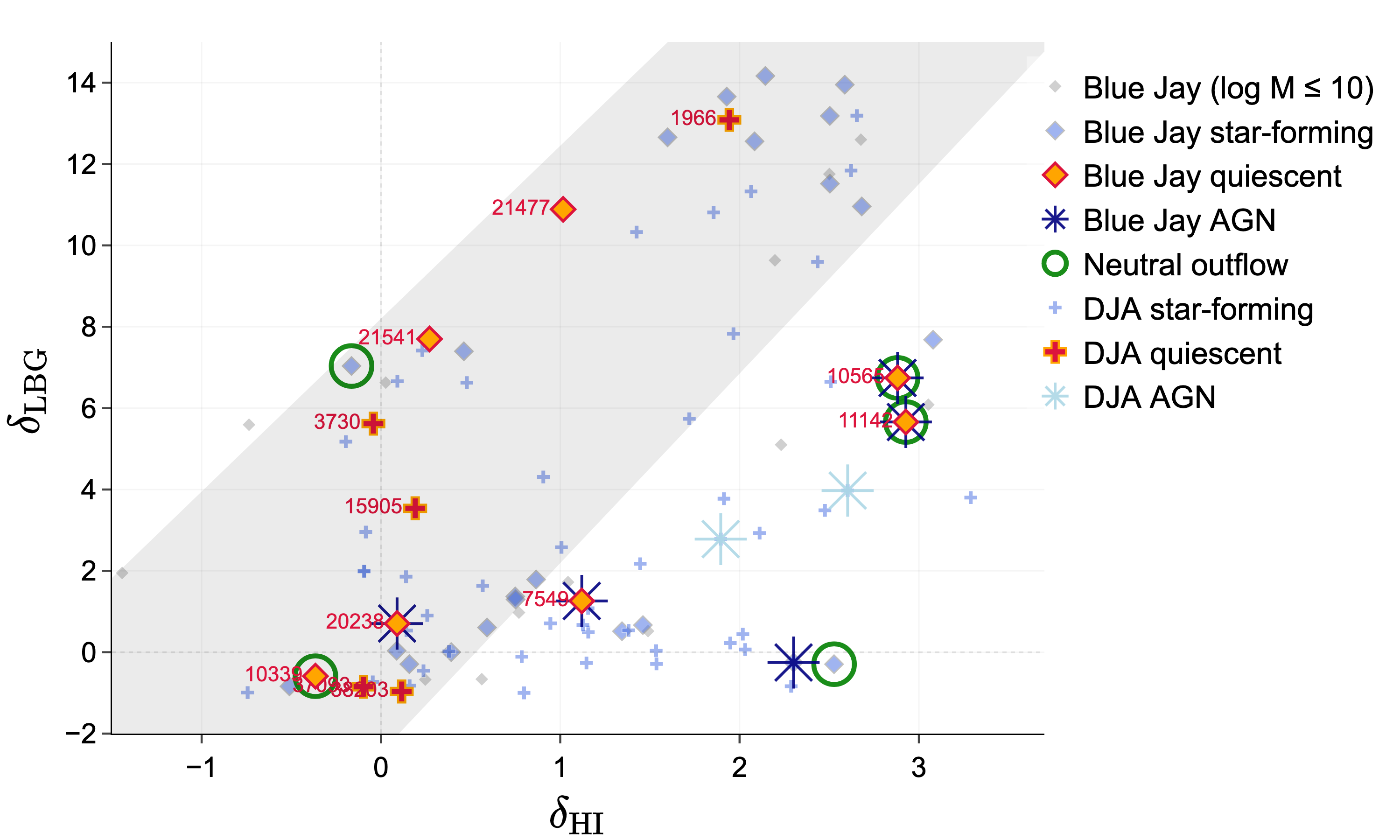}
    \caption{Number density $\delta_\mathrm{LBG}$ derived from LBGs, as a function of neutral-hydrogen density \dhi. Most of the sample follows a trend between these two measurements, qualitatively indicated by the gray band, to be compared to the trend between \dn\ and \dhi\ shown in Fig.~\ref{fig:dh_dn}. Gas-rich, galaxy-poor galaxies, such as 11142 and 10565, are found below this trend.}
    \label{fig:3D}
\end{figure*}

\end{appendix}

\end{document}